# Subterahertz collective dynamics of polar vortices

Qian Li[1,11†], Vladimir A. Stoica[1,2†], Marek Paściak[3], Yi Zhu[1], Yakun Yuan[2], Tiannan Yang[2], Margaret R. McCarter[4], Sujit Das[4], Ajay K. Yadav[4], Suji Park[7], Cheng Dai[2], Hyeon Jun Lee[8], Youngjun Ahn[8], Samuel D. Marks[8], Shukai Yu[2], Christelle Kadlec[3], Takahiro Sato[9], Matthias C. Hoffmann[9], Matthieu Chollet[9], Michael E. Kozina[9], Silke Nelson[9], Diling Zhu[9], Donald A. Walko[1], Aaron M. Lindenberg[7,9,10], Paul G. Evans[8], Long-Qing Chen[2], Ramamoorthy Ramesh[4,5,6], Lane W. Martin[4,6], Venkatraman Gopalan[2], John W. Freeland[1], Jirka Hlinka[3], Haidan Wen[1*]

[1] Advanced Photon Source, Argonne National Laboratory, Lemont, IL 60439, USA.

[2] Department of Materials Science and Engineering, The Pennsylvania State University, University Park, PA 16802, USA.

[3] Institute of Physics of the Czech Academy of Sciences, Na Slovance 2, 182 21 Prague 8, Czech Republic.

[4] Department of Materials Science and Engineering, University of California, Berkeley, CA 94720, USA.

[5] Department of Physics, University of California, Berkeley, Berkeley, CA 94720, USA

[6] Materials Sciences Division, Lawrence Berkeley National Laboratory, Berkeley, CA 94720, USA

[7] SIMES, SLAC National Accelerator Laboratory, Menlo Park, CA 94025, USA.

[8] Department of Materials Science and Engineering, University of Wisconsin-Madison, Madison, Wisconsin 53706, USA.

[9] Linac Coherent Light Source, SLAC National Accelerator Laboratory, Menlo Park, CA 94025, USA.

[10] Department of Materials Science and Engineering, Stanford University, Stanford, CA 94025, USA.

[11] School of Materials Science and Engineering, Tsinghua University, Beijing, 100084, China

*Correspondence to: wen@anl.gov

[†]equal contributions.

**The collective dynamics of topological structures[1–6] have been of great interest from both fundamental and applied perspectives. For example, the studies of dynamical properties of magnetic vortices and skyrmions[3,4] not only deepened the understanding of many-body physics but also led to potential applications in data processing and storage[7]. Topological structures constructed from electrical polarization rather than spin have recently been realized in ferroelectric superlattices[5,6], promising for ultrafast electric-field control of topological orders. However, little is known about the dynamics of such complex extended nanostructures which in turn underlies their functionalities. Using terahertz-field excitation and femtosecond x-ray diffraction measurements, we observe ultrafast collective polarization dynamics that are unique to polar vortices, with orders of magnitude higher frequencies and smaller lateral size than those of experimentally realized magnetic vortices[3]. A previously unseen soft mode, hereafter referred to as a vortexon, emerges as transient arrays of nanoscale circular patterns of atomic displacements, which reverse their vorticity on picosecond time scales. Its frequency is significantly reduced at a critical strain, indicating a condensation of structural dynamics. First-principles-based atomistic calculations and phase-field modeling reveal the microscopic atomic arrangements and frequencies of the vortex modes. The discovery of subterahertz collective dynamics in polar vortices opens up opportunities for applications of electric-field driven data processing in topological structures with ultrahigh speed and density.**

Precisely engineered $(PbTiO_3)_n/(SrTiO_3)_n$ oxide superlattices provide a controllable platform to host novel phenomena such as negative capacitance[8,9] and light-induced supercrystals[10] as well as unique polarization topologies including vortex[5] and skyrmion[6] structures. In comparison with their counterparts in magnetic systems[3,4,11], the building elements of these extended nanostructures are electric polarizations, which could lead to collective dynamics and properties that do not exist in magnetic systems: the direct interaction of polarization with electric fields allows electric-field excitation of their dynamics on ultrafast time scales, and the intrinsic coupling of polarization with the lattice offers strain tunability. These advantages have motivated the research of ferroelectrics for next-generation electronics. For example, nanoscale ferroelectric domain walls have been envisaged for ultrahigh-speed microelectronics and telecommunications[12–16] and host collective dynamics at terahertz (THz) frequencies[17–20], exceeding the dynamics of magnetic vortices and skyrmions that typically occurs in a few gigahertz range[7,21]. However, such terahertz dynamics in topological polar structures has not been experimentally demonstrated.

Collective dynamics also hold the key to understanding the many-body interactions in these newly discovered topological polar structures where long-range interactions beyond nearest neighbors are crucial[22]. The dynamics of soft modes that exhibit a reduction of the mode frequency at a critical point is important to understanding the thermodynamics of phase transitions[23], the access of hidden ferroelectricity far from equilibrium[24], and the condensation of metastable polar phases[25]. However, whether topological structures host new soft modes and how they behave on ultrafast time scales are unanswered questions for elucidating the fundamental physics and exploring the novel properties in these emergent nanostructures.

We focus our study on polar vortices[5], a prototypical topological structure of nanoscale ferroelectrics. The unique connectivity of electric polarization ($\mathbf{P}$) that continuously rotates around a core with non-zero curl ($\nabla \times \mathbf{P}$) in each vortex supercell can give rise to a series of new collective modes. Response spectra computed by dynamical phase-field modeling summarize a series of modes consistent with what we experimentally observed in the vortex structure (orange arrows,

Fig. 1a). These modes are distinct from optical phonons in bulk ferroelectrics[23], acoustic modes in superlattices[26], and chiral phonons in a hexagonal lattice[27]. Besides high-frequency modes in the range of 0.3-0.4 THz, we identified a low-frequency *vortexon* mode with a tunable characteristic frequency $\omega$ ("V" mode in Fig. 1a) by thermally induced strain in the low-frequency regime. The atomistic simulations show that this mode emerges as a transient nanoscale circular pattern of collective atomic motion (purple circles, Fig. 1b). Its time-dependent vorticity $(\nabla \times \boldsymbol{u})$ of atomic displacements ($\boldsymbol{u}$) oscillates on picosecond time scales, interweaving with polar vortices (magenta circles). A zoom-in snapshot shows the nanoscale collective circular patterns spanning many unit cells with a lateral size of ~6 nm, significantly smaller than the typical sub-micrometer size of magnetic vortices.

Ultrafast THz pump, hard x-ray diffraction probe measurements at the Linac Coherent Light Source (Fig. 1c and Methods) unveiled these collective modes in 100-nm-thick $(PbTiO_3)_{16}/(SrTiO_3)_{16}$ superlattices grown on $DyScO_3$ substrates. This technique circumvents the challenges of detecting incoherent modes in heterogeneous systems because the coherent THz field acts as an impulsive excitation to synchronize the polarization dynamics[28–31] while the x-ray diffraction selectively probes the vortex and conventional $a_1/a_2$-domain (FE) structures[32]. These structures coexist in the sample (Methods) but their diffraction signals can be probed separately in reciprocal space due to distinct lattice parameters (Fig. 2a and Extended Data Fig. 1a). The polarization direction of the applied THz field with respect to the crystalline axes was determined by the diffraction geometry and sample orientation (Extended Data Fig. 1b).

The structural response of the polar vortices upon THz excitation can be categorized into two frequency regimes. The high-frequency regime features multiple modes in the range of 0.3-0.4 THz, observed by probing selected Bragg peaks (Fig. 2a). For example, the diffraction intensities of the 023 peak and its satellites were modulated at the same frequency but with opposite phases (Fig. 2b). This indicated that the weakening or enhancement of the in-plane vortex order redistributed the diffraction intensity from the satellites to the main peak or vice versa. The THz field directly excited the superlattice layer rather than the substrate since there was no measurable structural response of the latter. The Fourier analysis of the intensity oscillations revealed the frequency components peaked at 0.34 and 0.38 THz (the orange arrows in Fig. 2c). To simultaneously capture the response of the conventional ferroelectric (FE) domain structure, we monitored the 113 reflections of both the FE superlattice peak and the vortex satellites with identical excitation conditions (Fig. 2b). The FE domain structure[32] exhibited a spectral fingerprint at 0.22 THz (the blue arrow in Fig. 2c and Extended Data Fig. 2), in analogy to the previously predicted inhomogeneous phason mode[18]. Its frequency is different from that of the vortex structure, suggesting that the collective dynamics of FE and vortex structures are uniquely associated with their microscopic polarization configurations. The diffraction signals monitored across the 023 FE Bragg peak oscillated in phase (Extended Data Fig. 2c,d), showing that the Bragg peak was modulated in the overall peak intensity as the structure factor of the FE domain changed, rather than shifting or broadening of the peak. The magnitude of the effects was proportional to the peak THz field, in agreement with a field-driven mechanism rather than a parabolic dependence due to THz-induced heating (Extended Data Fig. 2b).

Complementary microscopic pictures of the collective modes were obtained through atomistic and dynamical phase-field modeling (Methods and Supplementary Notes S1 and S2), with the former seeking the equilibrium eigenmodes based on first-principles-derived ionic potentials[18] and the latter addressing the driven dynamics of the polarization order parameter[33,34]. The atomistic model revealed multiple prominent eigenmodes in the same high-frequency regime (Extended Data Fig. 3). Microscopic dynamical properties are illustrated using the calculated 0.30 THz mode as an example because its calculated dielectric response is relatively large and likely to be observed. Snapshots of its polarization vectors are shown at the equilibrium state ($t < 0$) and the maximum ($t = \tau/4$) of the atomic displacement during a sinusoidal oscillation (Fig. 2d and Supplementary Video S1). In this mode, the central positions of the polar vortices do not change along the *z*-axis while the dominant lead-cation displacement is along the *x*-axis (Fig. 2e). To illustrate the diversity of the dynamics in this frequency range, complex configurations of polarizations and atomic displacements of additional modes are shown in Extended Data Fig. 4. Their infrared activity corresponds well with the observed THz absorption (Supplementary Note S7).

The low-frequency regime features the vortexon mode at 0.08 THz at room temperature, which was observed from the diffraction intensity change of the 004-vortex peak (Fig. 3a). Remarkably, this unique mode was highly sensitive to temperature. An increase in sample temperature from 293 to 388 K resulted in significant hardening of this mode from 0.08 to 0.23 THz (Fig. 3a,b), in contrast to any known modes in bulk or strained $PbTiO_3$ and $SrTiO_3$. The amplitude of the oscillation decreased as the sample temperature increased, correlated well with the weakening of the vortex order[32]. The intensities of the δ04-satellite peaks with a small in-plane scattering component $\delta = \pm 0.059$ Å$^{-1}$, exhibit strong high-frequency oscillations at 0.34 and 0.38 THz, consistent with the measurements of the 023 and 113 peaks. However, these frequencies did not change as a function of temperature. At higher temperatures, closer to the vortex-ordering temperature of 473 K[32], the vortex order was further reduced so that the x-ray diffraction intensity was too weak to measure. The oscillation amplitudes of all modes were reduced after 36 ps, which was the time for the completion of lattice expansion (Extended Data Fig. 2g,h). Possible causes for this amplitude reduction may be related to the leakage of the mechanical energy into the substrate[35].

The observation of the 0.08 THz mode in the vortex structure is corroborated by a low-frequency mode which is well separated from high-frequency ones in the atomistic model. Examining its polarization dynamics, we found that the centers of a pair of polar vortices shift in opposite directions along the *z*-axis (Fig. 3c, Supplementary Video S2), different from the gyration of magnetic vortex cores[3]. The polarization change $\Delta\mathbf{P}$ forms a wavy structure as a result of collective motions of lead, titanium, and oxygen atoms. The corresponding lead-cation displacements are linear within the unit cell, in contrast to the circular atomic motion of chiral phonons[27,36]. They collectively form a circular pattern, producing a transient non-zero curl of atomic displacement $\nabla \times \boldsymbol{u}(\boldsymbol{t})$ (Fig. 3d), hence the name vortexon. This circular pattern of atomic displacement gives rise to a transient angular momentum with respect to the vortexon cores (Supplementary Note S1B). Titanium and oxygen ionic motions follow similar patterns (Extended Data Fig. 4). The maximum displacement occurs at locations with a large curl of polarization, suggesting that the pre-existing polarization gradient within the supercell facilitates the structural response to the THz field. As a result, the vortexon cores (center of violet circles in Fig. 1b) are spatially shifted by half of the period with respect to the core of the equilibrium polar-vortex structure (center of the magenta circles in Fig. 1b). The orientation of the THz field with respect to the crystalline axes can influence the amplitude but not the frequency of this mode. Experimentally, when the THz field is aligned perpendicular to the x-axis, the vortex response is stronger than that in the case when the THz field is parallel with the x-axis (Extended Data Fig. 5a, Supplementary Note S5), which is different from the prediction by the simplified models (Supplementary Note S6).

We identified the dominant role of thermally induced strain in producing the observed large tunability of the vortexon mode. The variation of the strain due to a temperature rise of 95 K, confirmed by x-ray diffraction measurements (Extended Data Fig. 7a,b), amounts to a lattice expansion of 0.1% along the x-axis. We then theoretically explored the effect of strain (ε) on the eigenmodes. The zero strain was defined specifically in the atomistic and phase-field models (Supplementary Note S1 and S2). These two independent theoretical approaches produced the same trends (Fig. 4). Numerous modes lying in the 0.3-0.4 THz range were insensitive to strain variation while a distinct low-frequency mode can be effectively tuned by strain, in agreement with the observed in-plane strain variations in the range of -0.1 to 0.04% due to the temperature changes from 293 to 388 K. The softening of the frequency towards a critical strain $\varepsilon_c$ (-0.3% in the atomistic model and -0.2% in phase-field simulation) is phenomenologically similar to phonon softening close to a structural phase transition. Indeed, the atomistic models show that the polar vortices go through a symmetry-breaking transition from symmetric to staggered vortex-core pairs (Fig. 4a, bottom inset) across the critical strain, different from temperature-induced order-disorder vortex phase transition[32]. The equilibrium state consists of the staggered vortices in which the vortex cores are offset oppositely along the *z*-axis, away from the central positions of the vortex supercell. The small-amplitude oscillation around this equilibrium position results in the modulation of diffraction intensity at the fundamental frequencies of the modes, rather than their second harmonics. An analytical model (Supplementary Note S3) further shows the motion of vortex cores can be parametrized as the vertical displacement $\xi$ of vortex cores (Fig. 3c) away from their equilibrium positions, to which specular reflections such as the 004 peak are particularly sensitive. The analytical model also reveals the anharmonic nature of the vortexon (Supplementary Note S3), which opens the route for controlling topological structures via nonlinear phononics[37].

To understand the structural response in the time domain, we employed dynamical phase-field simulations to calculate the time-dependent ferroelectric polarization and construct the corresponding atomic positions (Supplementary Note S2, Extended Data Fig. 8). The simulated polarization dynamics captured the eigenmodes obtained from the atomistic calculations (Supplementary Videos S3 and S4). The simulated Bragg-peak intensities as a function of time reproduced the main experimental observations of vortexon modes (Fig. 4b). The subtle difference between the simulated and experimental data for the high-frequency modes is due to the simplifications of the models (Supplementary Note S6). We estimate that ~4% intensity modulation of the 004 vortex superlattice peak corresponds to maximum lead displacements of ~ 5 picometers (Extended Data Fig. 6e).

To distinguish observed collective dynamics from the conventional superlattice acoustic modes[26,38], we performed a control experiment on the same sample wherein we replaced the THz excitation by a 400 nm, 100 fs optical excitation[39]. The intensity modulations of the 004 peak exhibited the characteristic frequencies of 0.43 and 0.56 THz, corresponding to coherent acoustic waves propagating along the out-of-plane direction at the longitudinal and transverse sound speed in the superlattice, respectively (Supplementary Note S4, Extended Data Fig. 9). Therefore, the comparison of the responses upon 400-nm and THz excitations unambiguously distinguishes collective modes of polar vortices from the conventional superlattice acoustic modes.

In summary, we experimentally identified a new set of collective modes that arise from the unique connectivity of electric polarization and its coupling with the lattice in topological ferroelectric nanostructures. The strain-tunable vortexon mode exhibited collective atomic motions in circular patterns. The combined experimental and theoretical studies unveiled their eigenfrequencies and eigenvectors. This level of understanding is critical to exploring new physics and applications of topological structures. For example, the condensation of the vortexon mode at a critical strain could provide a new avenue of phonon engineering in complex oxides. The dynamical phase-field simulation further predicts the tunability of vortexon mode by external electric field (Extended Data Fig. 6c,d), offering effective control of the polar-vortex structure at room temperature in nanodevices. These modes have orders of magnitude higher intrinsic frequency and smaller size than their magnetic counterparts, allowing direct electric-field control of their dynamics for high-speed, high-density data processing and storage.

## Main References:

**Acknowledgments:** We acknowledge the discussions with Mariano Trigo, Di Xiao, Zijian Hong, Igor Luk'yanchuk and Valerii M. Vinokur. This work was primarily supported by the U.S. Department of Energy, Office of Science, Basic Energy Sciences, Materials Sciences and Engineering Division: experimental design, data collection, data analysis, and part of simulations by Q.L. and H.W. were supported under the DOE Early Career Award; ultrafast measurements and sample synthesis by V.A.S., Y.Y., S.P., L.W.M., C.D., S.Y., A.L., L.-Q.C., V.G., J.W.F., and H.W. were supported under Award Number DE-SC-0012375; ancillary ultrafast x-ray measurements by H.L., S.M., Y.A., and P.E. were supported under Award Number DE-FG02-04ER46147. M.M., S.D., and R.R. acknowledge support for part of sample synthesis through the Quantum Materials program (KC 2202) funded by the U.S. Department of Energy, Office of Science, Basic Energy Sciences, Materials Sciences Division under contract No. DE-AC02-05-CH11231. J.H., M.P. were supported by the Czech Science Foundation (project no. 19-28594X) and acknowledge the access to computing facilities owned by parties and projects contributing to the National Grid Infrastructure MetaCentrum, provided under the program No. Cesnet LM2015042. T.Y. and L.-Q.C. acknowledge the partial support from the US Department of Energy, Office of Science, Basic Energy Sciences, under Award Number DE-SC0020145 as part of the Computational Materials Sciences Program and from NSF under Award DMR-1744213. Y. Z. and H.W. acknowledge the support by ANL-LDRD for preliminary x-ray measurements. Q.L. acknowledges support by the Basic Science Center Project of NSFC under grant no. 51788104 for completing phase-field simulations at Tsinghua University. S.M. acknowledges support from the Office of Science Graduate Student Research (SCGSR) program (DOE contract number DE-SC0014664) and from the UW-Madison Materials Research Science and Engineering Center (NSF DMR-1720415). H.L. acknowledges support by the National Research Foundation of Korea under grant 2017R1A6A3A11030959. Use of the Linac Coherent Light Source is supported by the US Department of Energy, Office of Science, Office of Basic Energy Sciences under contract no. DE-AC02-76SF00515. Use of the Advanced Photon Source is supported by the US Department of Energy, Office of Science, Office of Basic Energy Sciences under contract no. DE-AC02–06CH11357.

**Author contributions:** Q.L., V.S., Y.Y., M.M., S.P., H.L., Y.A., S.M., T.S., M.H., M.C., M.K., S.N., D.Z., A.L., P.E., V.G., J.F. and H.W. performed the experiment at the Linac Coherent Light Source. Y.Z., V.S., D.W., and H.W. performed preliminary experiments at the Advanced Photon Source. M.P. and J. H. performed atomistic modeling. Q.L., D.C., T.Y., and L.-Q.C. performed phase-field simulation. T.Y., L.-Q.C. and. J.H. developed the analytical model. S.D., M.M., A.Y., L.W. and R.R. prepared the samples. S.Y. and C. K. performed THz spectroscopy measurements.

Q. L. and H.W. wrote the manuscript with input from all authors. H.W. conceived and supervised the project.

**Competing interests:** Authors declare no competing interests.

**Additional information:**

**Extended data** is available for this paper at https://doi.org/...

**Supplementary information** is available for this paper at https://doi.org/...

**Reprints and permissions information** is available at http://www.nature.com/reprints.

**Correspondence and requests for materials** should be addressed to H.W.

**Main Figure**

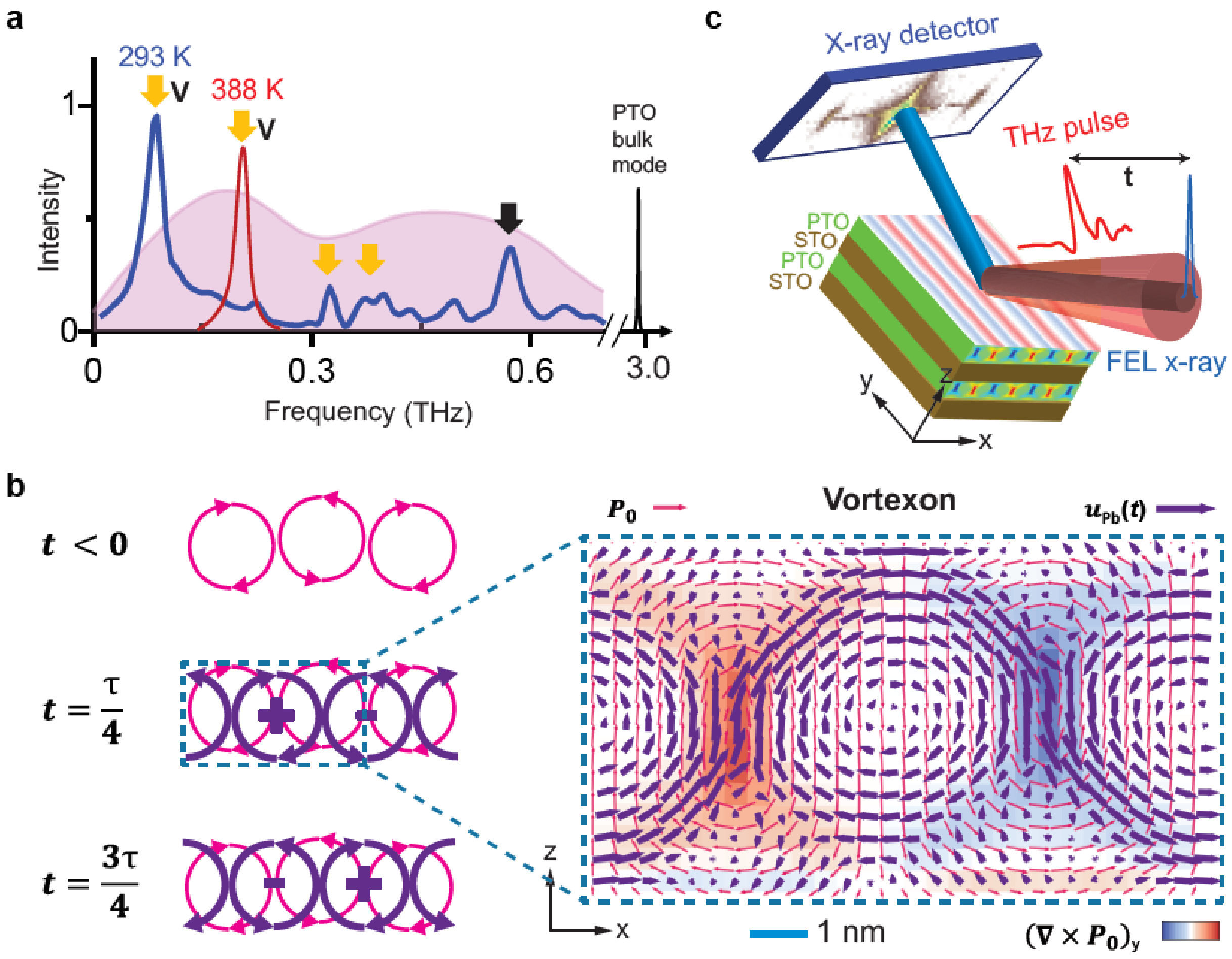

**Fig. 1 | Emergence of the collective dynamics of polar vortices and its experimental detection.** **a**, Fourier spectra (blue) of calculated time-dependent response of polar vortices upon THz-field excitation. The spectrum of the THz pulse is shown as a broad pink background. The collective modes of polar vortices (orange arrows) are shown as a separate set of modes with respect to the known superlattice acoustic modes (black arrow) and the soft mode of $PbTiO_3$ at room temperature. The vortexon mode marked by "V" shifts to higher frequency (red peak) as sample temperature increases. **b**, Emergence and evolution of the vortexon (atomic displacement vortices, purple circles) during its oscillation period τ, overlaid with the static polarization vortices (magenta circles). "+" and "-" label the signs of vortexon vorticity that reverse dynamically. A zoomed-in

view of the region of the dotted box is shown on the right with the calculated static polarization (magenta arrows) and lead-cation displacement (purple arrows) in each unit cell of the vortexon mode at $t = \frac{\tau}{4}$. **c,** Schematic of THz-pump and x-ray-diffraction-probe experiment at x-ray free-electron laser (FEL). The color stripes on the $(PbTiO_3)_{16}/(SrTiO_3)_{16}$ (PTO/STO) superlattice film represent in-plane vortex orders with opposite polarization vorticity.

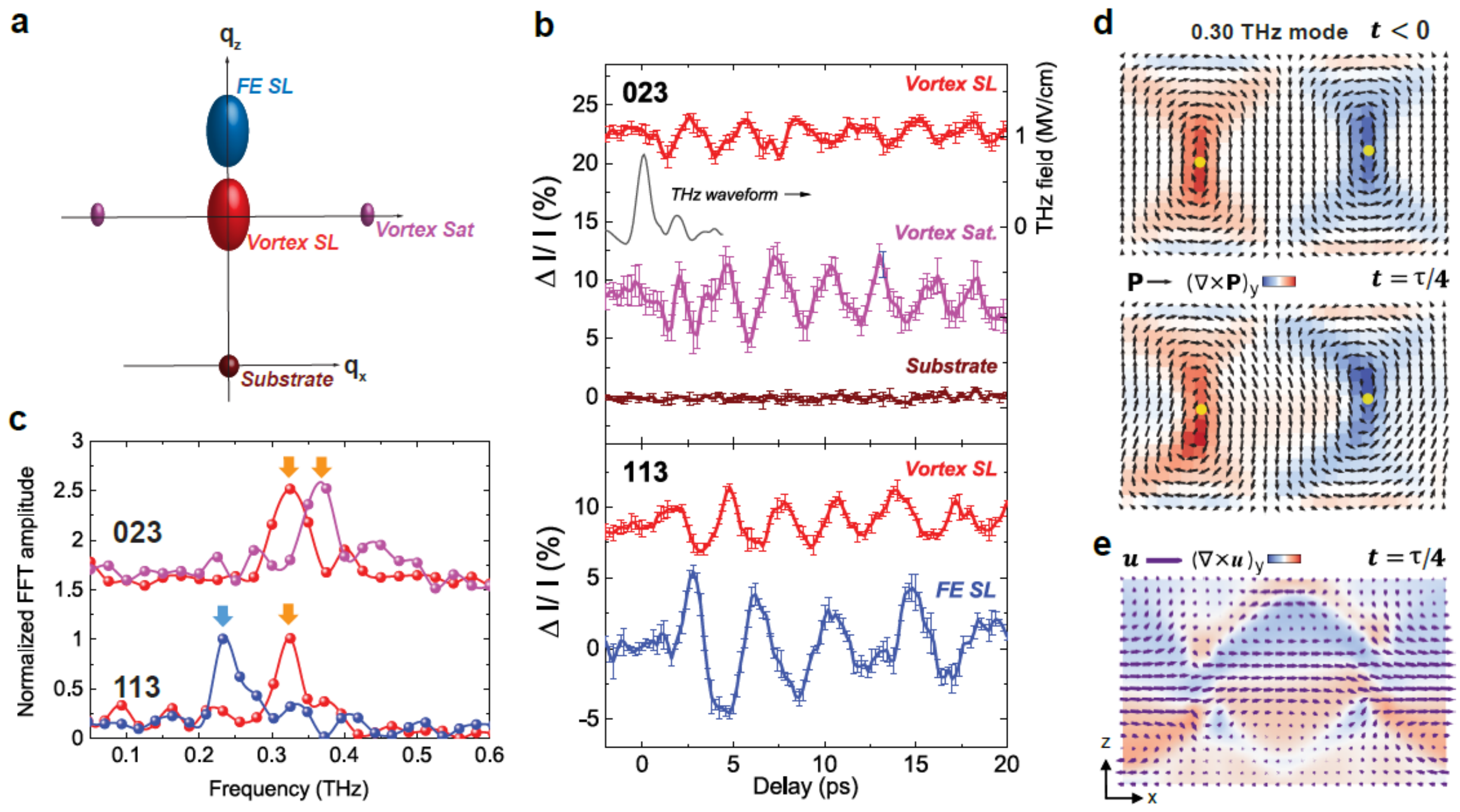

**Fig. 2 | High-frequency collective modes. a,** Schematic of the probed Bragg peaks in the $q_x$-$q_z$ plane of the reciprocal space around 023, 113, 004 substrate peaks. Substrate: $DyScO_3$ peak; SL: superlattice peak; Sat: satellite peak; FE: ferroelectric $a_1/a_2$ structure. **b**, Diffraction intensity of the Bragg peaks as indicated by the same color scheme as in (a) as a function of delay, with the measured THz excitation pulse overlaid. The error bars show the standard errors. **c**, The corresponding Fourier spectra with marked vortex modes (orange arrows) and FE modes (blue arrow). The curves in (b) and (c) are offset vertically for clarity. **d**, Polarization (**P**) of a calculated 0.30 THz vortex mode at the equilibrium state ($t < 0$) and the sinusoidal maximum ($t = \tau/4$), with associated vorticity (color). **e**, Lead-cation displacements ($\boldsymbol{u}$) and associated vorticity (color) at $t = \tau/4$. The structure distortion is enhanced for better visibility.

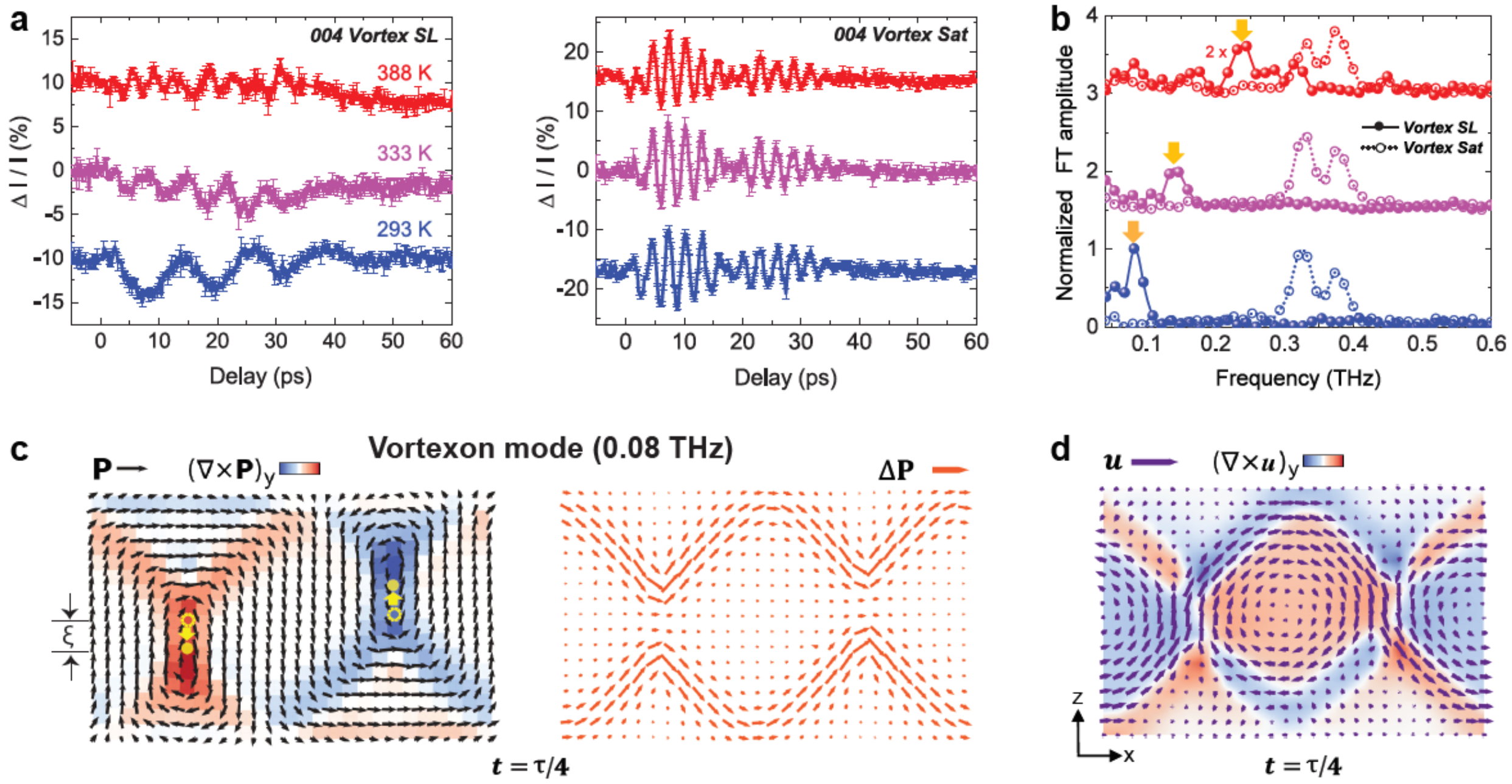

**Fig. 3 | Tunable vortexon mode. a**, Diffraction intensity of 004-superlattice (SL) and δ04-satellite peaks of vortex structures as a function of delay at various temperatures. The error bars show the standard errors. **b**, Corresponding Fourier spectra of the intensity oscillations of the 004 SL peak (solid) and δ04-satellite peaks (open circles), normalized to the spectra at 293 K. The orange arrows indicate the tunable vortexon mode. The curves in (a) and (b) are offset vertically for clarity. **c**, Calculated polarization (**P**) and the change of polarization (**ΔP)** of the vortexon at the sinusoidal maximum (t = τ/4). The yellow arrows indicate the shift of vortex cores (solid dots) from the equilibrium positions (open yellow circles) by a distance $\xi$ comparing with the case of t < 0 shown in Fig. 2c. **d**, Lead-cation displacements ($\boldsymbol{u}$) and associated vorticity (color) at t = τ/4. The vector lengths of $\boldsymbol{u}$ and the shift $\xi$ are enlarged arbitrarily and 2.5 times for better visibility, respectively.

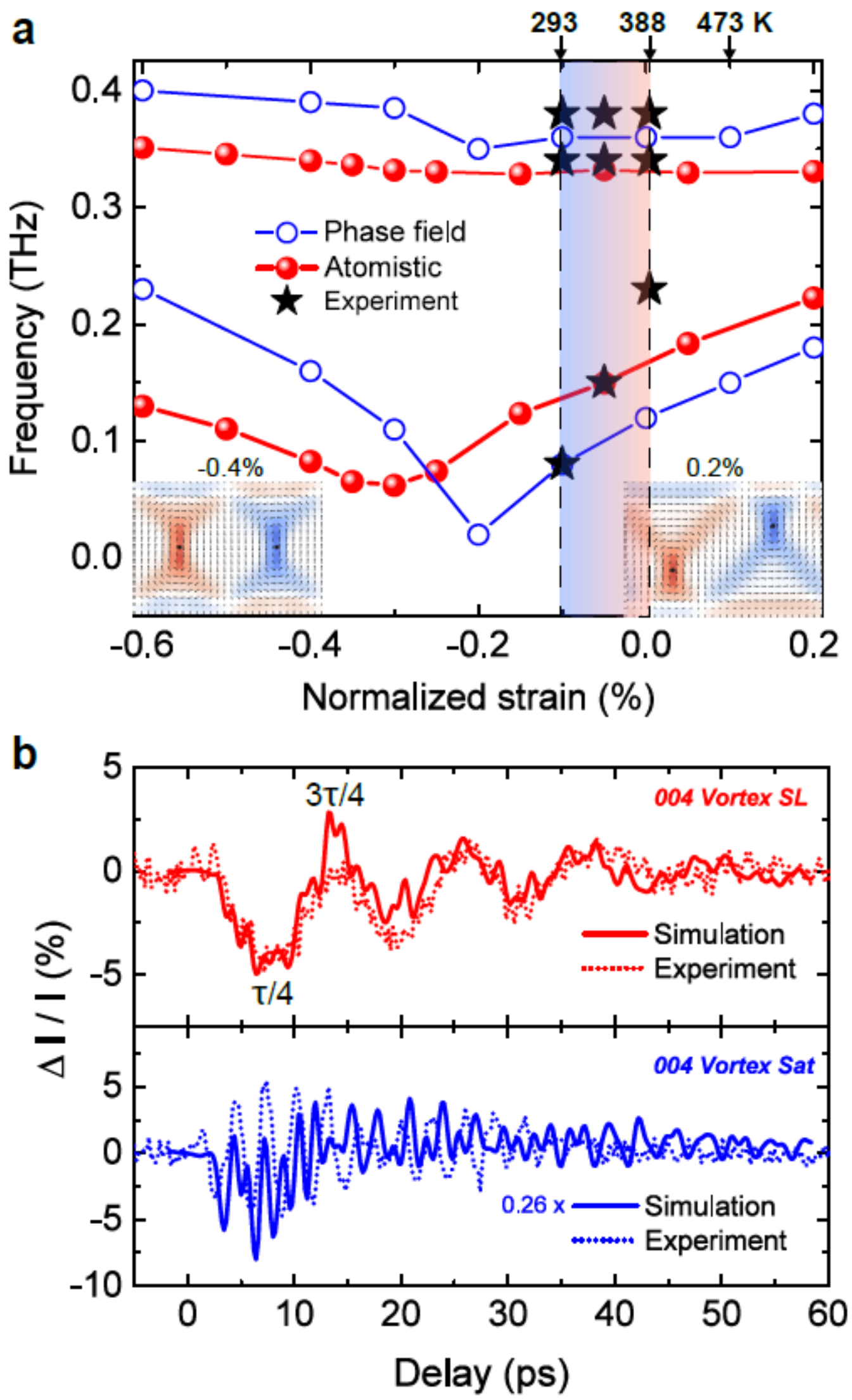

**Fig. 4 | Atomistic calculation and dynamic phase-field simulation. a**, Calculated frequencies of the collective vortex modes as a function of normalized strain in comparison with experimental data. The vertical dashed lines indicate the strain at 293 K and 388 K, below the vortex melting temperature of 473 K. The schematics of polar vortex pairs at the bottom show the calculated equilibrium states at -0.4 and 0.2% strain with the same legend as in Fig. 2(d). **b**, Simulated (solid) and measured (dot) diffraction intensity of the 004 Bragg peaks as a function of time at 293 K. The minimum and maximum of intensity change of 004 vortex SL peak at the delay of τ/4 and 3τ/4 correspond to two configurations shown in Fig. 1b.

**Methods:**

***Terahertz or optical-pump, ultrafast x-ray diffraction measurements.*** The measurements were performed at the X-ray Pump Probe beamline of the Linac Coherent Light Source (LCLS)[40]. The LCLS delivered horizontally polarized x-ray pulses with a pulse duration of 40 fs at a repetition rate of 120 Hz. Hard x-ray pulses at 9.5 keV with 0.1% bandwidth were selected using a diamond (111) monochromator and focused to ~200×200 $\mu m^2$ beam size with beryllium compound refractive lenses. The x-ray beam was attenuated to ~$10^9$ photons per pulse to avoid sample damage. The sample was mounted on a multi-circle goniometer for both vertical and horizontal scattering geometries. The temperature of the sample stage is controllable within the range of 293 K to 388 K in ambient air using a homebuilt resistive heater. Scattered x-rays were captured shot-by-shot using a high dynamic-range area detector (Jungfrau 1M), positioned at a radial distance of 600−750 mm from the sample.

A Ti: Sapphire femtosecond laser system synchronized to the free-electron laser was used to generate 100 fs, 800 nm ultrafast laser pulses with up to 20 mJ pulse energy, which were then converted to single-cycle terahertz pulses using the pulse-front-tilt technique in a $LiNbO_3$ crystal as detailed in Ref.[41]. The THz pulses were vertically polarized with a maximum peak electric field of ~800 kV $cm^{-1}$, as calibrated by the electro-optic (EO) sampling with a GaP crystal. The THz beam was focused on the sample and collinearly aligned with the primary x-ray beam using a 90-degree off-axis parabolic mirror. In a separate optical pump setup, 400 nm optical pump pulses were derived by frequency doubling of the 800 nm laser pulses with an incident fluence of 3 $mJ\cdot cm^{-2}$. The spatiotemporal overlap between optical and x-ray pulses and between optical and THz pulses were achieved using "t0-finder" as detailed in Ref.[42].

X-ray scattering patterns around selected diffraction peaks were acquired as a function of delay between the THz/optical pump and x-ray probe pulses with a temporal resolution of ~50 fs, as limited by the timing jitter with the correction by the "timing" tool. A fast measurement scheme was adopted in which the delay was continuously back-and-forth scanned within time delays up to ~40 ps using a mechanical stage; an electronic delay mechanism was applied in combination for longer delays[43]. The raw detector images were grouped into laser on/off datasets (50% laser firing event ratio), normalized against the intensity monitor $I_0$ (rejecting shots with the $I_0$ lying outside a predefined range), and binned to 100 - 200 fs temporal slots as needed. Typically, ~500-1000 effective shots were accumulated for each time sampling data point.

***Sample growth and characterization.*** The $(PbTiO_3)_{16}/(SrTiO_3)_{16}$ superlattice samples on $DyScO_3$ (DSO) substrate were fabricated using reflection high-energy electron diffraction (RHEED) - assisted pulsed laser deposition[5]. Structural characterization and preliminary time-resolved measurements of the samples were carried out using synchrotron-based x-ray diffraction (XRD) at beamlines 33-ID-C and 7-ID-C of the Advanced Photon Source, respectively. Three-dimensional reciprocal space maps (RSMs) characterized the structural properties of polar vortex and $a_1/a_2$ ferroelectric (FE) structures in the superlattices, which were analyzed to determine the strain of both phases as a function of temperature (Extended Data Fig. 7). The FE structures exhibit conventional $a_1/a_2$ stripe domains with 1.6% smaller out-of-plane lattice constant than the vortex structures and thus they can be separately probed by x-ray diffraction. These structures have average widths of ~200 nm and the intermixed regions do not contribute significantly to the measured Bragg diffraction intensity.

***Atomistic modeling.*** The all-atom calculations made use of a parametrization of atomic interactions in the form of the shell model[44]. Therein each atom is represented by a core and a shell, connected by an anharmonic spring, which is designed to mimic atomic polarizability. The potential energy of the system is calculated as a sum of long-range Coulomb interactions between all particles except for cores and shells of the same atom, short-range interactions between shells of oxygen-cation and oxygen-oxygen pairs, and the anharmonic spring interaction for the same-atom core-shell pairs. $PbTiO_3$ (PTO) parameters were taken from Sepliarsky and Cohen[45] who fitted their model to the results of first-principle calculations. This type of PTO potential has been shown to correctly predict vortex structures in a layered material[46]. The $SrTiO_3$ (STO) parameters were adjusted from Sepliarsky et al.[44] to match the atomic charges of PTO, which was necessary to avoid charged interfaces in the layered PTO/STO superstructure. The refined STO parameters are given in Supplementary Table S1. For the Ti and O atoms within an interface atomic layer, averages of the PTO and STO parameters were taken. Program GULP[47] was used for the structure optimization and lattice dynamics calculations, the details of which are presented in Supplementary Note S1.

***Dynamical phase-field calculation.*** Based on the Landau-Ginzburg-Devonshire (LGD) theory, the $PbTiO_3/SrTiO_3$ thin-film superlattice system can be described by a free-energy function $F$ in the following form:

$$F = \alpha_{ij}P_iP_j + \alpha_{ijkl}P_iP_jP_kP_l + \alpha_{ijklmn}P_iP_jP_kP_lP_mP_n + \frac{1}{2}g_{ijkl}P_{i,j}P_{k,l} + \frac{1}{2}c_{ijkl}\varepsilon_{ij}\varepsilon_{kl} - q_{ijkl}\varepsilon_{ij}P_kP_l - \frac{1}{2}\kappa_b E_iE_i - E_i^{total}P_i \quad \text{(Eq. 1)}$$

where all repeating subscripts imply summation over the Cartesian coordinate $x_i$ ($i$ = 1, 2 and 3) and ‘,$i$’ denotes the partial derivative with respect to $x_i$ (that is, $\partial/\partial x_i$); $P_i$ is one component of the polarization vector **P**, $e_{ij}$ is elastic strain, and $E_i^{total} = E_i + E_i^{THz}$ is the total electric field including the effective THz field $E_i^{THz}$ and internal field $E_i$; $\alpha$’s are Landau coefficients (up to the sixth order for PTO and the fourth order for STO), $g_{ijkl}$ are polarization gradient energy coefficients related to the domain wall energy, $c_{ijkl}$ are elastic stiffness tensors, $q_{ijkl}$ are electrostrictive coupling coefficients and $\kappa_b$ (isotropic) is the background dielectric permittivity.

We implemented the phase-field model using the finite-element method[33]. The model included three sets of field variables: mechanical displacement **u** = [$u_x$, $u_y$, $u_z$], electric potential [$\varphi$] so that $E_i = -\varphi_{,i}$, and polarization vector **P** = [$P_x$, $P_y$, $P_z$],; other physical quantities in Eq. 1 can be derived from these variables, for example, $E_x = -\varphi_{,x}$ and $\varepsilon_{xz} = 0.5(u_{x,z} + u_{z,x})$. The Cartesian model coordinates correspond to the pseudocubic crystallographic axes of the samples, as shown in Fig. 1c. For computing efficiency, we studied a thin-slab 3D superlattice model with an *xyz* dimension of 108×0.8×152.4 $nm^3$, consisting of 8 alternating thin-film layers of 6.4 nm PTO and 6.4 nm STO (approximately 16 unit-cell thickness for each layer) together with a 50 nm substrate layer of DSO. The film was assumed to be fully constrained to the substrate; the displacement of the bottom surface (*xy*) of DSO was fixed while the top film surface was set mechanically free. In-plane continuity periodic boundary conditions were imposed for the two opposing pairs of *yz* and *xz* surfaces. Short-circuit electrical boundary conditions (i.e., $\varphi = 0$) were applied for both top and bottom film surfaces. The mechanical strain state of the entire system was calculated with reference to the cubic-phase PTO, assuming the lattice constants of cubic PTO (3.955 Å), STO (3.905 Å) and DSO (3.951 and 3.945 Å along the *x* and *y* axes, respectively). The model geometry was discretized into a mesh of 0.8 nm cube elements with the quadratic Lagrange shape function.

The spatiotemporal dynamics of the superlattice system is governed by the following set of coupled equations of motion[34]:

$$\mu \ddot{P}_i + \gamma \dot{P}_i = -\delta F / \delta P_i \qquad \text{(Eq.2)}$$

$$\rho \ddot{u}_i + \beta \rho \dot{u}_i = \left( \frac{\partial F}{\partial \varepsilon_{ij}} \right)_{,j} \qquad \text{(Eq. 3)}$$

$$0 = \partial F/\partial \varphi \quad \text{(Eq. 4)}$$

in which $\mu$ is the effective polarization mass density, $\gamma$ is polarization damping coefficient, $\rho$ is mass density, and $\beta$ is the elastic damping coefficient. Eq. 2 is a second-order time-dependent LGD equation that explicitly controls the dynamics of the polarization order parameter, while Eq. 3 explicitly controls the elastodynamics (yet they are coupled via the electrostriction effect). In formulating Eq. 4, we imply that there is no complex electrodynamics associated with electron/hole carrier transport since the intrinsic carrier concentration in the system can be neglected and low-energy THz excitation (versus 400 nm optical pump) does not populate free carriers in this system. Note that Eq. 3 governs both the superlattice film and substrate layers whereas Eq. 2 and 4 are only defined for the film layers.

Typically, we first nullified the mass inertial effects by setting both the $\mu$ and $\rho$ as zero and generated a quasi-steady state vortex configuration from an initial random state consisting of small polarization values ($< 0.001P_0$, $P_0 = 0.7$ C/m$^2$). We then turned on the mass parameters and equilibrated the model for ~40 −100 ps prior to executing excitation runs. To simulate the THz excitation, we added a time-dependent E-field pulse $E_i^{THz}(t)$ to the internal E-field which had a realistic THz waveform as measured from the electro-optical sampling and combined *x*/*y*/*z* components according to a specific measurement condition. The peak amplitude of the applied E-field was 100−200 kV/cm, taking into account the dielectric screening by the PTO/STO film due to the background permittivity, and was finely tuned to match the measured x-ray scattering signals. The generalized-alpha time stepping method was adopted to solve the Eqs.2-4, where the maximum step was limited to 0.1 ps to eliminate high-frequency numerical errors. Further details of the dynamical phase-field calculation and the simulation of time-dependent x-ray diffraction intensity can be found in Supplementary Note S2. The response spectra, which displays the time-dependent changes of averaged polarization $\langle P^2 \rangle$ in Fig. 1a, were calculated upon THz excitation with a peak electric field of $1 \times 10^7$ V/m along the *x*-axis using the parameters listed in Supplementary Table S2.

**Methods References:**

**Data availability:** The data that support the findings of this study are available from the corresponding author upon request.

**Code availability:** The code that support the findings of this study are available from the corresponding author upon request.

## Extended Data Figures

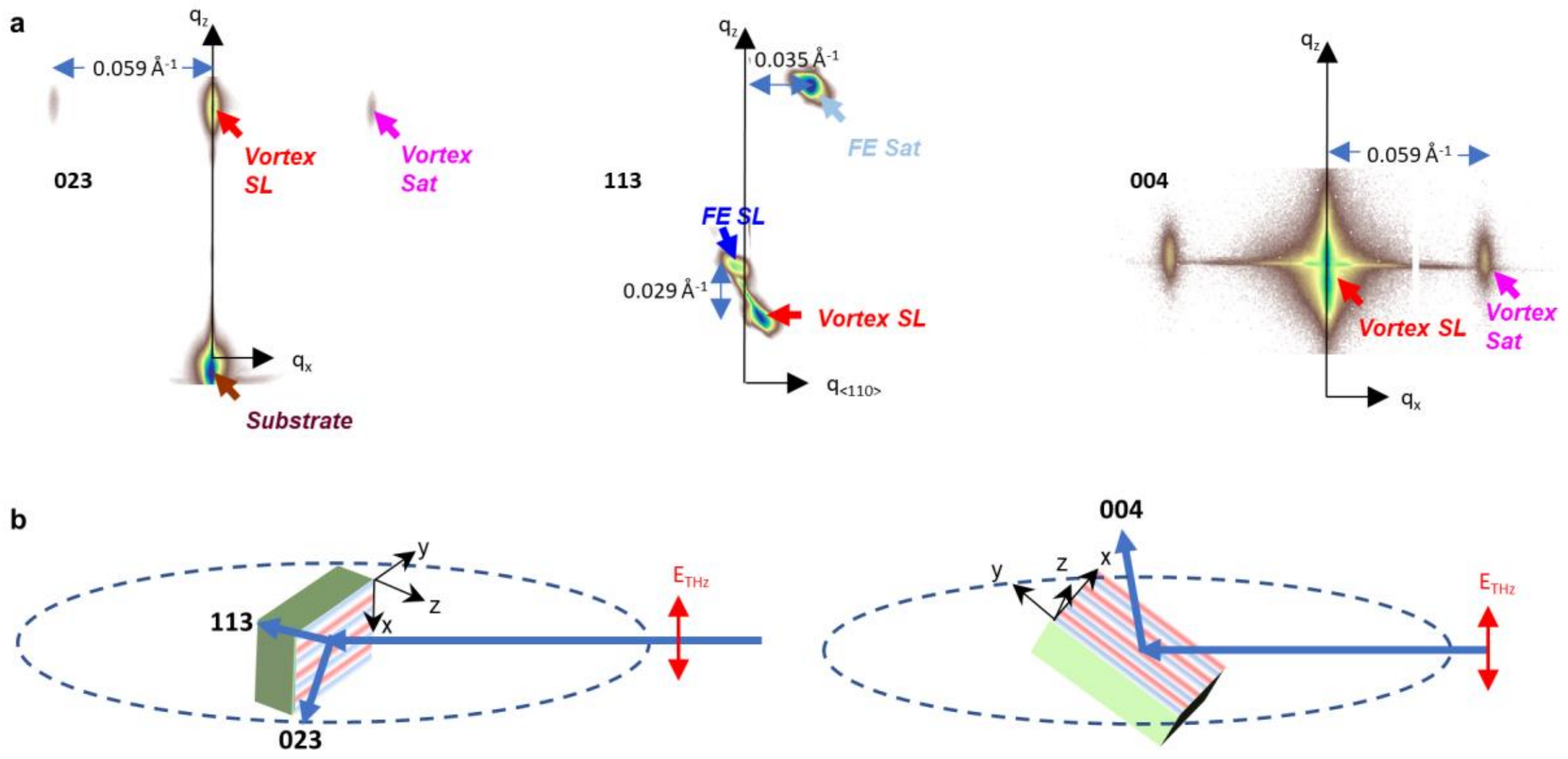

**Extended Data Fig. 1 | Diffraction patterns of probed Bragg peaks and the corresponding diffraction geometry**. **a**, Diffraction patterns of 023, 113 and 004 peaks in logarithmic scale as recorded by a two-dimensional x-ray area detector. The projected q-axis labels indicate the approximate directions in the reciprocal space. **b**, Schematics of diffraction geometry for probing 023, 113, and 004 peaks. The dashed circle represents the horizontal plane. The red and blue color stripes represent the vortex structure with opposite vorticities. The THz field is polarized vertically.

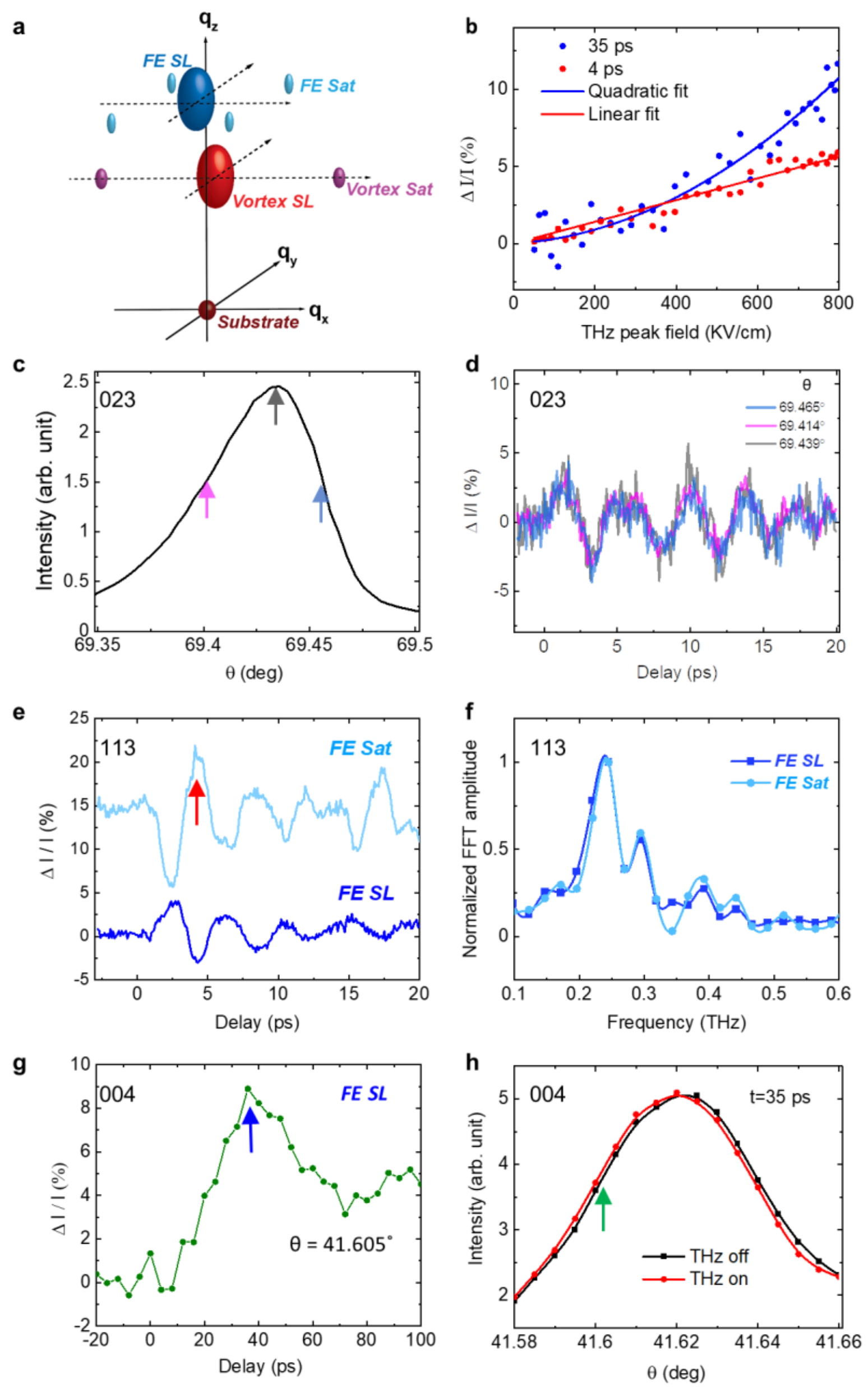

**Extended Data Fig. 2 | The response of the FE ($a_1/a_2$) structure upon THz field excitation**. **a**, 3D schematic of diffraction peaks around the 023 or 113 substrate peaks in the reciprocal space. The dashed arrows indicate the reciprocal axes through the centers of the FE or vortex SL peaks that are offset along the $q_y$-axis due to the lattice constant difference along the *y*-axis. Only two SL and their satellite peaks were shown for simplicity. **b**, Diffraction intensity of the 113 FE satellite peak (red curve) measured at the delay of 4 ps indicated by the red arrow in (e), and diffraction

intensity of the 004 FE SL peak (blue curve) measured at lower angle side and at the delay of 35 ps indicated by the blue arrow in (g), as a function of the applied THz peak field. Lines are linear and quadratic fits to the measured data, in agreement with the field-driven structure factor modulation and the THz-induced Bragg peak shift (heating), respectively. **c**, Rocking curve of 023 FE peak. **d**, Diffraction intensity as a function of delay measured at various incident angles which are indicated by the magenta, grey, and blue arrows in (c). **e,** Diffraction intensity of the 113 peak and its satellites as a function of delay. The curves are vertically offset for clarity. **f**, Fourier spectra of (e). **g**, Diffraction intensity as a function of delay measured at the lower angle side of the 004 peaks, indicated by the green arrow in (h). **h**, 004 rocking curves measured at the delay of 35 ps with and without THz excitation. The peak shift shows a strain of $2.5 \times 10^{-5}$, corresponding to a 2 K temperature rise as calibrated by the temperature-dependent x-ray diffraction measurements shown in Extended Data Fig. 7b.

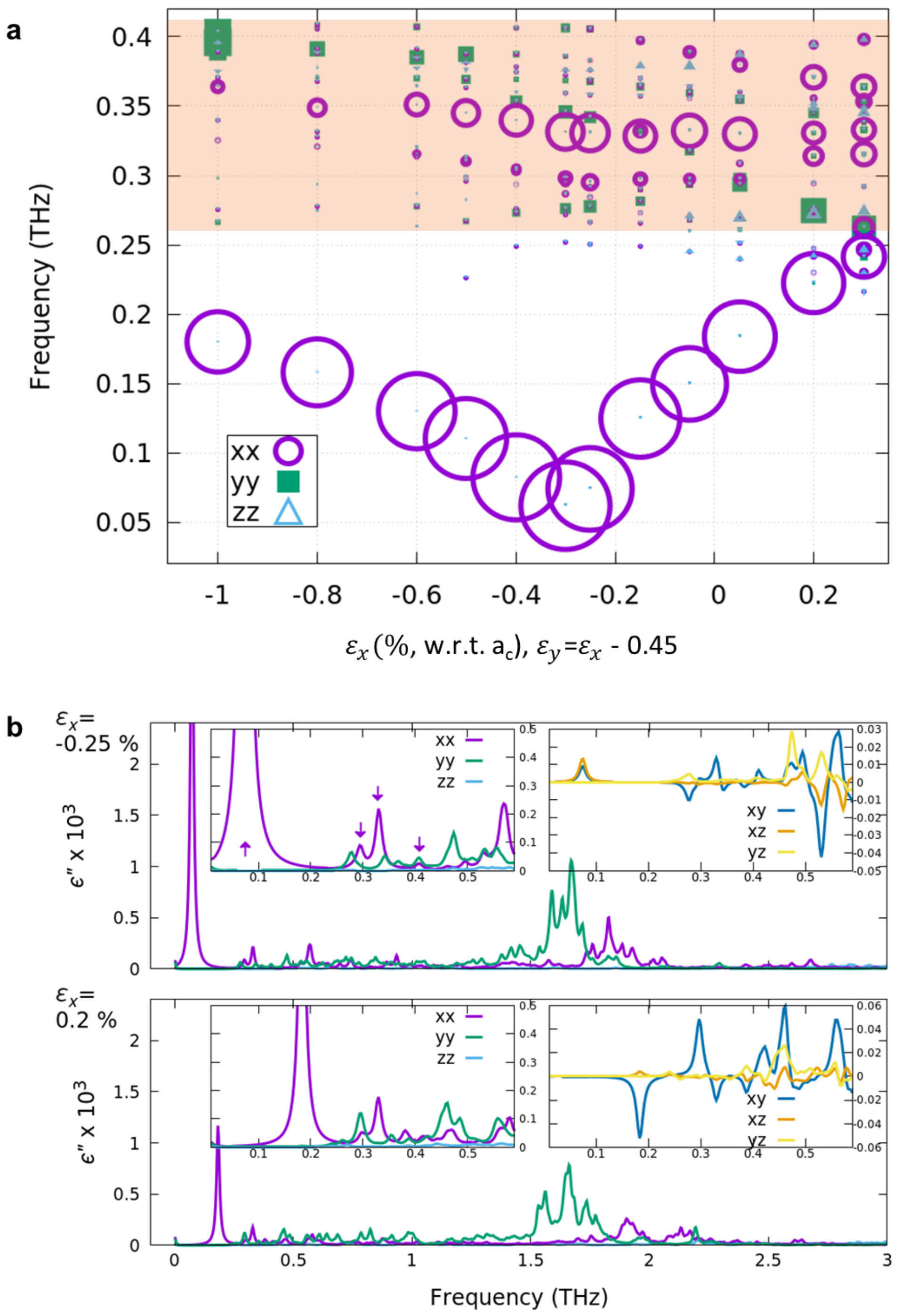

**Extended Data Fig. 3 | The dielectric properties of vortex structures as a function of strain and frequency from atomistic calculations. a**, Frequency-strain diagram with the diagonal oscillator strength ($S_{ii}$) of each mode reflected in the size of the markers [the area is proportional to log($S_{ii}$+1THz$^2$)], where i represents *x*, *y* or *z*. The dielectric strength of the m-th mode ($\Delta\epsilon_{ii,m}$)

is directly related to the oscillator strength via: $\Delta\epsilon_{ii,m}=S_{ii,m}/\omega_m^2$, where $\omega_m$ is the mode's frequency. The 15 lowest-frequency modes are considered; the *zz* signal is multiplied 5 times for better visibility. The orange area indicates the high-frequency regime. **b**, Frequency-dependent permittivity for two different strain values, $\varepsilon_x$=-0.25% and 0.2%. The imaginary part is presented with the main plots, while the left-hand insets show diagonal components $\epsilon''_{xx}$, $\epsilon''_{yy}$, $\epsilon''_{zz}$ and the right-hand insets present off-diagonal spectra $\epsilon''_{xy}$, $\epsilon''_{xz}$, $\epsilon''_{yz}$. The arrows indicate modes that are analyzed in the Extended data Fig. 4. All dielectric properties are calculated for the PTO layer only.

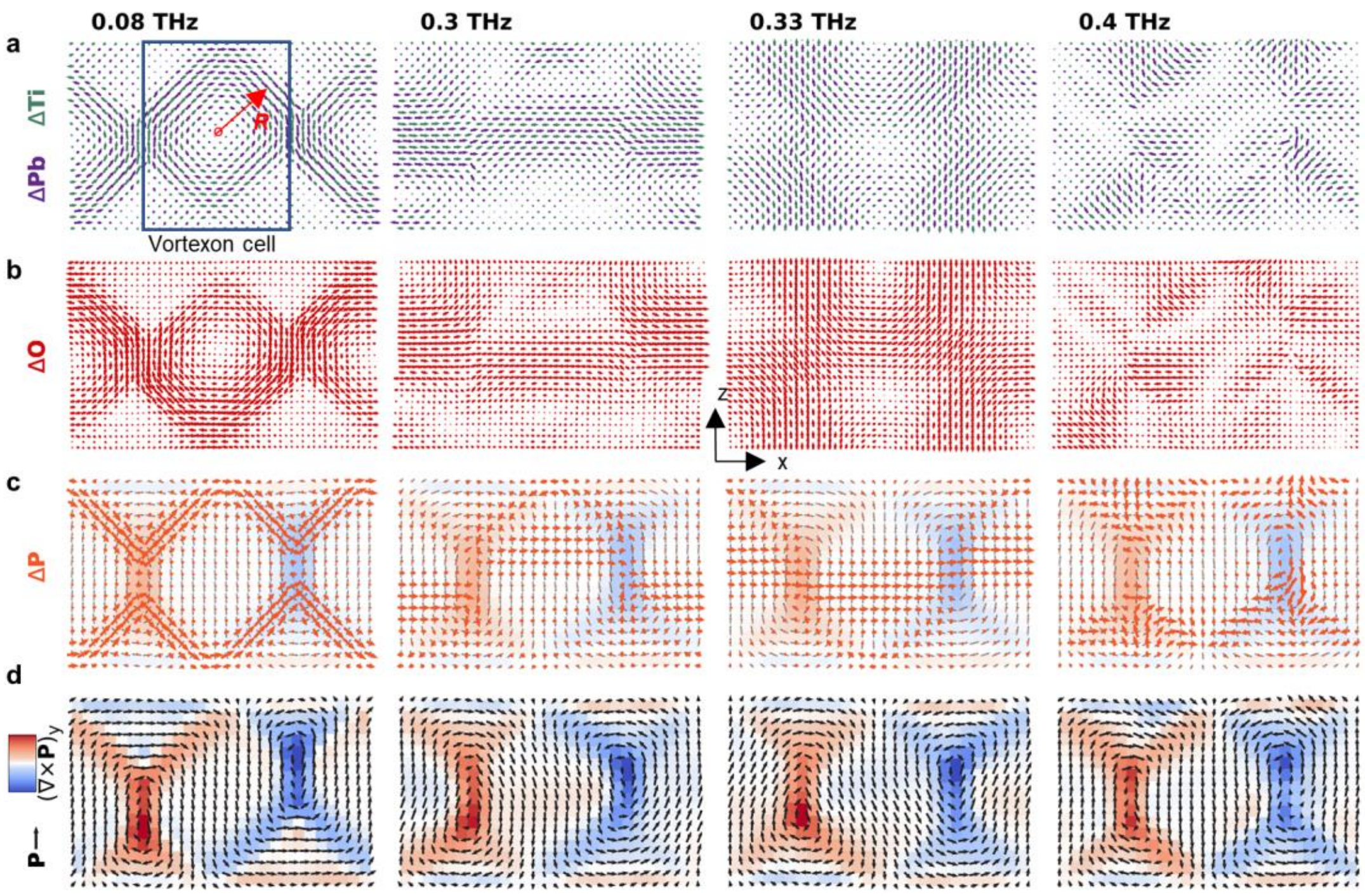

**Extended Data Fig. 4 | Decomposition of selected modes with strain $\varepsilon_x$ = -0.25 % in the atomistic model.** The modes are characterized by the highest oscillator strengths in the *x* direction (below the frequency of 0.41 THz, marked by the arrows in Extended Data Fig. 3b) and the first two of them are also described in the main text. Each column contains microscopic configurations at $t=\tau/4$ of **a**, Pb and Ti displacements; **b**, O displacements of the mode's eigenvectors; **c**, change in polarization caused by these atomic displacements (overlaid on a $\nabla \times \mathbf{P}$ map of the equilibrium structure); **d**, polarization patterns and the respective $\nabla \times \mathbf{P}$ map of a structure subjected to the mode's perturbation. The origin **O** and vector ***R*** in (a) are related to the calculation of the angular momentum of vortexon in the boxed region detailed in Supplementary Note S1B. The displacement vectors are scaled arbitrarily, while perturbed **P** configurations are calculated for maximum ionic displacements of 30 pm. We note that there are additional modes with the *y*-component of the oscillator strength that contribute to the diversity of the collective dynamics in the frequency range 0.3-0.4 THz.

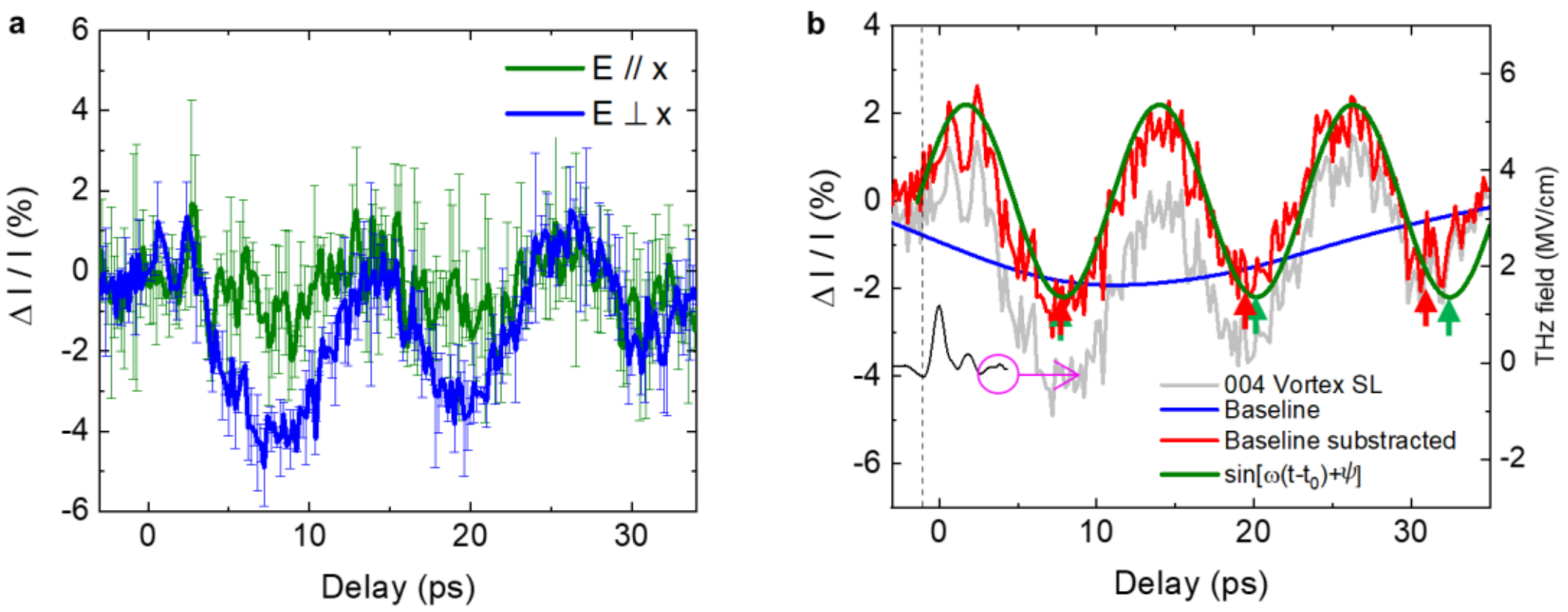

**Extended Data Fig. 5 | Polarization dependent response and analysis of the vortexon mode** **a**, Diffraction intensity of 004 vortex Bragg peak as a function of time when the THz field is applied parallel and perpendicular to the x-axis, respectively. **b**, Baseline-subtracted diffraction intensity of the 004 vortex superlattice peak (red curve) as a function delay. Raw data shown in grey color are from Fig. 4b. The green curve shows a sinusoidal wave (see Supplementary Note S5), with its minima (green arrows) deviating from the data (red arrows) at later time. The vertical dashed line indicates the time when the THz field turns on.

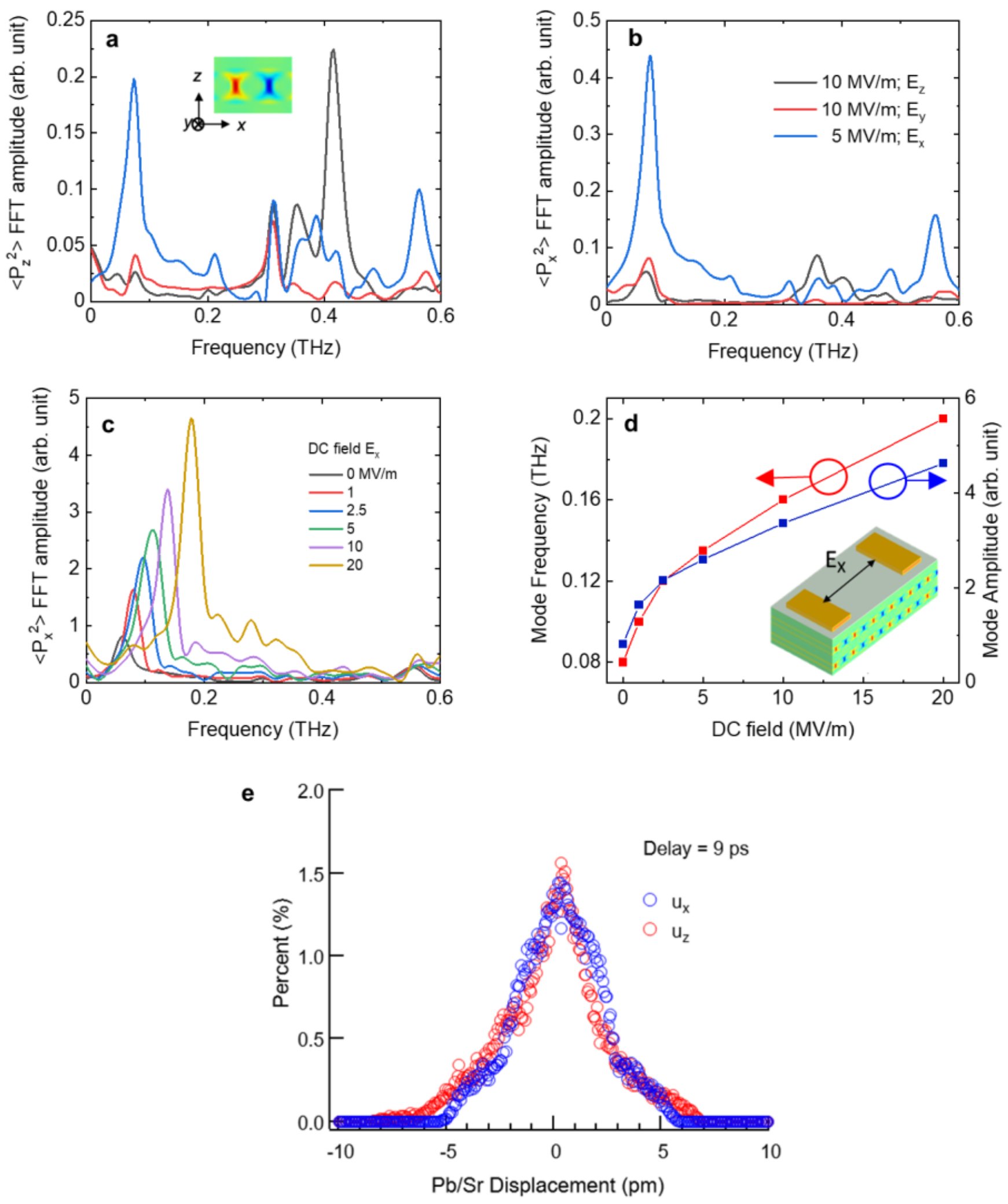

**Extended Data Fig. 6 | Dynamical phase-field simulation results.** Fourier transform of the simulated polarization dynamics, which is represented by spatially averaged (**a**) $\langle P_z^2 \rangle$ and (**b**) $\langle P_x^2 \rangle$ in the simulation box upon the application of the THz-field pulse aligned along the *x*-, *y*- or *z*-axis. **c**, Calculated frequency spectra of $\langle P_x^2 \rangle$ with various the direct-current (DC) electric field along the *x*-axis. **d**, Frequency and amplitude of the vortexon mode as a function of the applied electric field ($E_x$). Such an in-plane electric field can be applied using a pair of coplanar electrodes, as schematically shown in (**d**) inset. **e**, Histograms of the Pb and Sr ionic displacement ($\boldsymbol{u}$) away from its equilibrium positions along the *x* and *z* axes, corresponding to the maximum diffraction intensity change of 4% of 004 SL peak at a delay of 9 ps in the simulation.

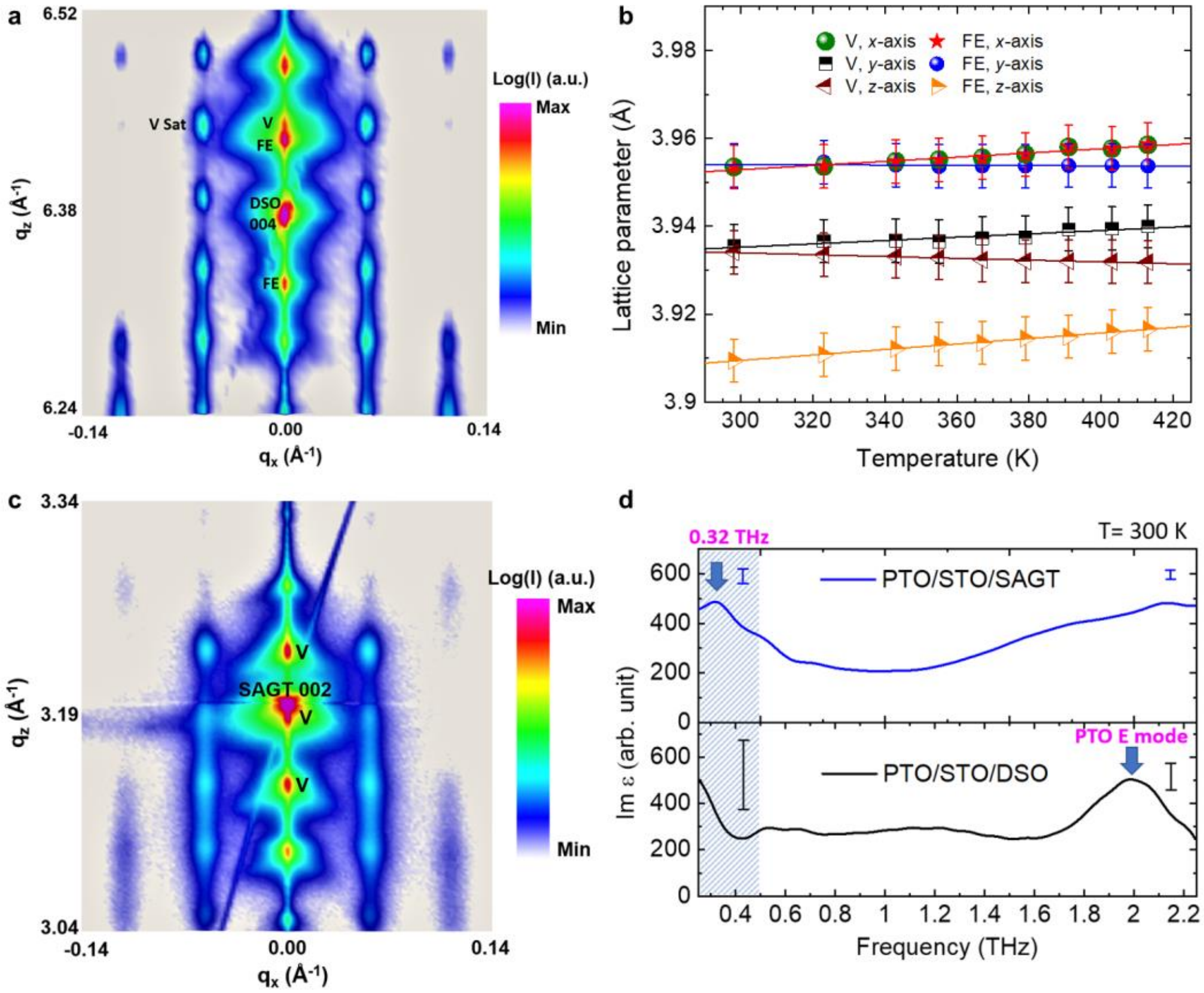

**Extended Data Fig. 7 | Equilibrium characterizations of $(PbTiO_3)_{16}/(SrTiO_3)_{16}$ superlattice.** **a**, Room-temperature reciprocal space map (RSM) of the superlattice grown on $DyScO_3$ (DSO) substrate cutting through the substrate specular 004 peak, which shows vortex (V) and FE superlattice diffraction peaks in the $q_x$-$q_z$ plane. The satellites (V Sat) due to the in-plane vortex ordering appear along $q_x$ direction around the corresponding superlattice (V) peaks. **b**, The measured average lattice parameters with linear fits for the vortex and FE structures as a function of temperature. **c**, Room-temperature RSM of the superlattice grown on $Sr_2Al_{0.3}Ga_{0.7}TaO_6$ (SAGT) substrate cutting through the substrate specular 002 peak, which shows the vortex SL diffraction peaks and satellites. **d**, Room-temperature imaginary dielectric constant extracted from the THz time-domain absorption spectroscopy measurements of the superlattices grown on DSO and SAGT substrates. The error bars represent the estimated experimental errors in the frequency ranges of shaded and unshaded regimes, respectively.

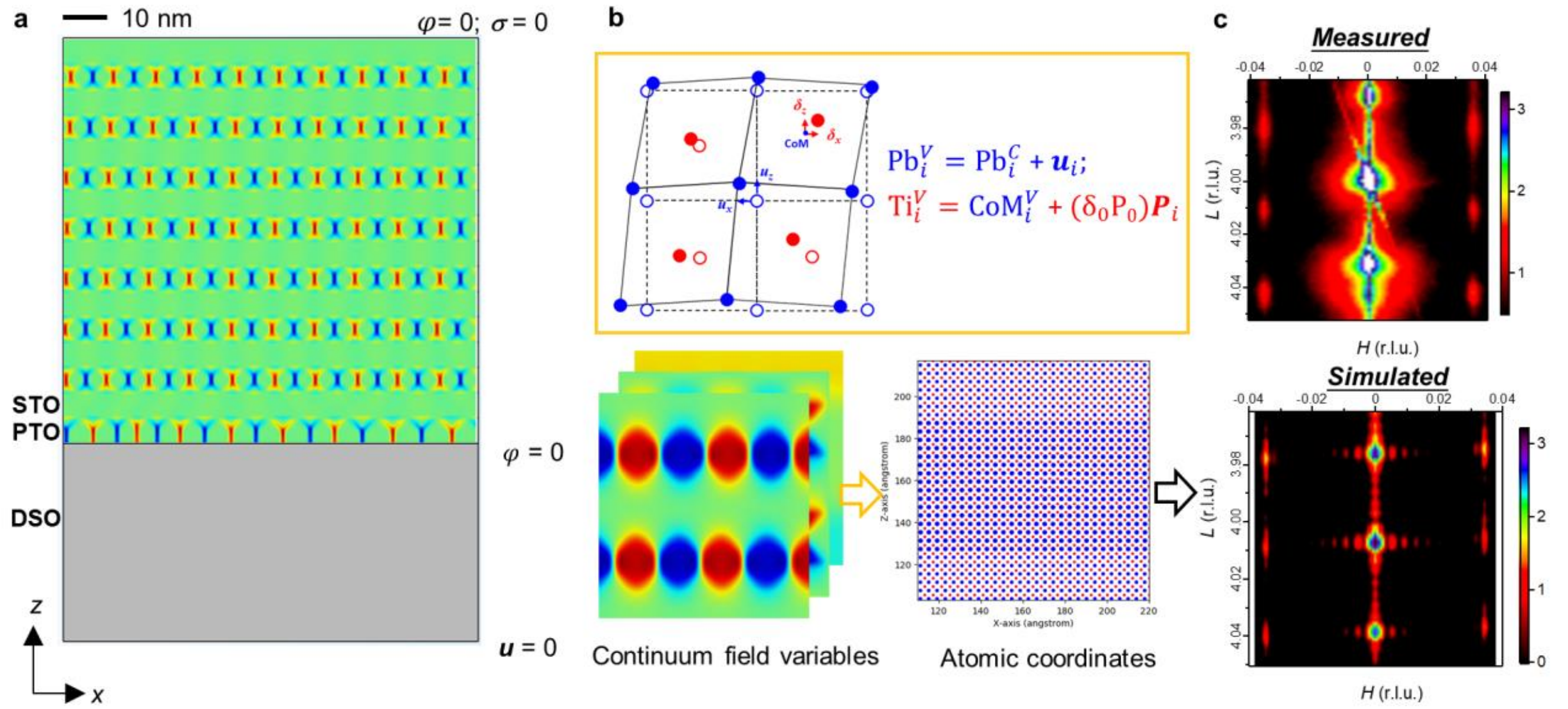

**Extended Data Fig. 8 | X-ray diffraction simulations based on dynamical phase-field modeling. a**, Model geometry of the $PbTiO_3/SrTiO_3$ (PTO/STO) superlattice film and the $DyScO_3$ (DSO) substrate. The color scale denotes the vorticity of the polarization vectors. The boundary conditions (Supplementary Note S2) are marked on the right side. **b**, Schematics for the method used to map the phase-field modeling output to atomistic configurations, based on which kinematic x-ray scattering intensities are calculated (Supplementary Note S2). **c**, Measured and simulated *H0L*-cut (along the $q_x$-$q_z$ plane) reciprocal space maps near the 004 reflection with the intensity shown in a logarithmic color scale. The fringes in the simulated RSM are due to the finite lateral size of the simulation box.

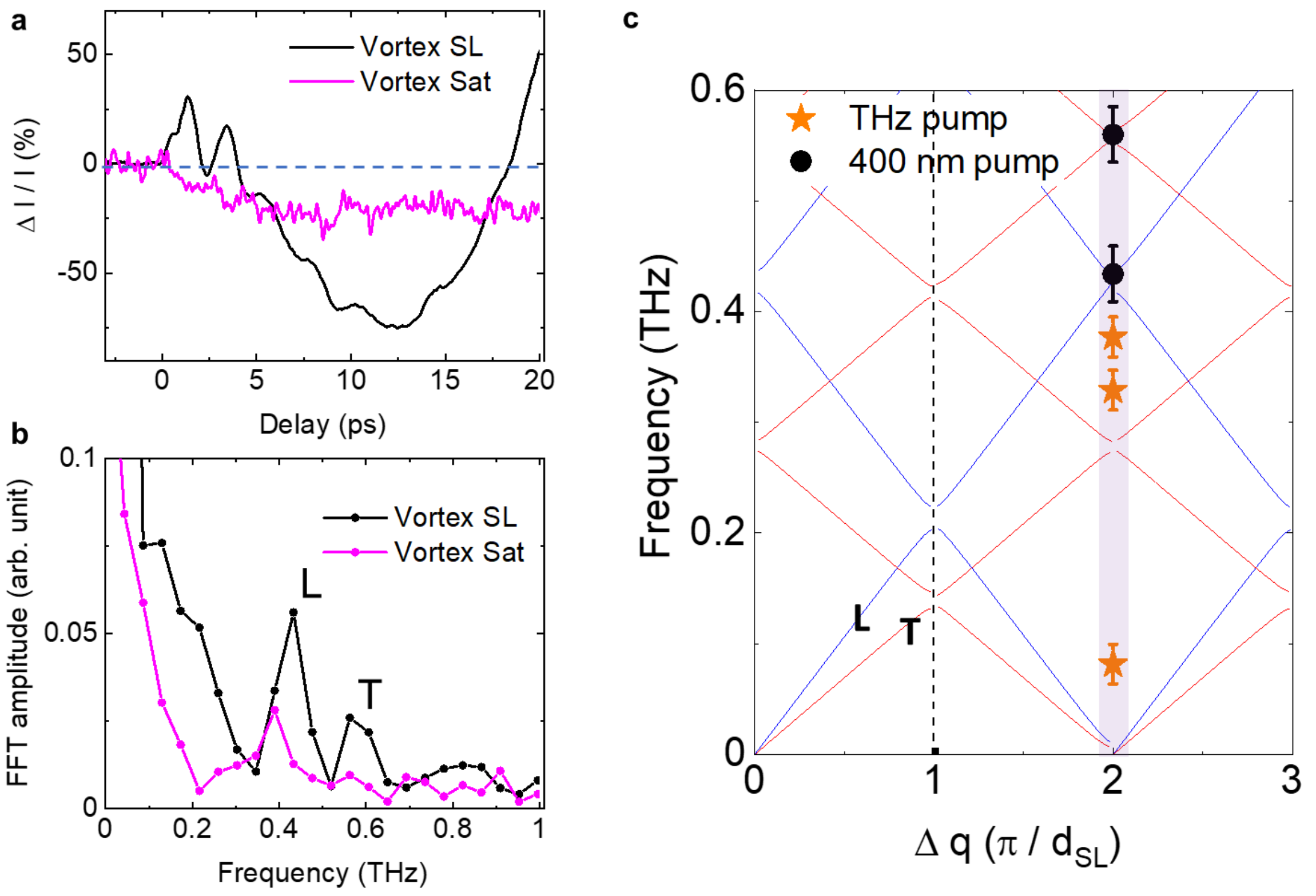

**Extended Data Fig. 9 | Structural dynamics excited by 400 nm optical pulses. a**, Normalized changes of 004-vortex peak intensity as a function delay. **b**, Fourier spectra of (**a**). “L” and “T” represent the acoustic waves that travel along the out-of-plane direction with the longitudinal and transverse sound speeds, respectively. **c**, Calculated phonon dispersion curves. The vertical dashed line indicates the boundary of the folded Brillouin zones. The shaded region corresponds to the scattering vectors probed in the experiment. The solid circles show the measured modes upon 400 nm excitation, while the stars show the measured modes upon THz excitation. $d_{SL}$ is the superlattice periodicity. The error bars show the full width at half maximum of the peaks in the Fourier spectra.

**Supplementary Information for**

# Subterahertz collective dynamics of polar vortices

**Authors:** Qian Li[1,11†], Vladimir A. Stoica[1,2†], Marek Paściak[3], Yi Zhu[1], Yakun Yuan[2], Tiannan Yang[2], Margaret R. McCarter[4], Sujit Das[4], Ajay K. Yadav[4], Suji Park[7], Cheng Dai[2], Hyeon Jun Lee[8], Youngjun Ahn[8], Samuel D. Marks[8], Shukai Yu[2], Christelle Kadlec[3], Takahiro Sato[9], Matthias C. Hoffmann[9], Matthieu Chollet[9], Michael E. Kozina[9], Silke Nelson[9], Diling Zhu[9], Donald A. Walko[1], Aaron M. Lindenberg[7,9,10], Paul G. Evans[8], Long-Qing Chen[2], Ramamoorthy Ramesh[4,5,6], Lane W. Martin[4,6], Venkatraman Gopalan[2], John W. Freeland[1], Jirka Hlinka[3], Haidan Wen[1*]

**Affiliations:**

1 Advanced Photon Source, Argonne National Laboratory, Lemont, IL 60439, USA.

2 Department of Materials Science and Engineering, The Pennsylvania State University, University Park, PA 16802, USA.

3 Institute of Physics of the Czech Academy of Sciences, Na Slovance 2, 182 21 Prague 8, Czech Republic.

4 Department of Materials Science and Engineering, University of California, Berkeley, CA 94720, USA.

5 Department of Physics, University of California, Berkeley, Berkeley, CA 94720, USA

6 Materials Sciences Division, Lawrence Berkeley National Laboratory, Berkeley, CA 94720, USA

7 SIMES, SLAC National Accelerator Laboratory, Menlo Park, CA 94025, USA.

[8] Department of Materials Science and Engineering, University of Wisconsin-Madison, Madison, Wisconsin 53706, USA.

[9] Linac Coherent Light Source, SLAC National Accelerator Laboratory, Menlo Park, CA 94025, USA.

[10] Department of Materials Science and Engineering, Stanford University, Stanford, CA 94025, USA.

[11] School of Materials Science and Engineering, Tsinghua University, Beijing, 100084, China

*Correspondence to: wen@anl.gov

†equal contributions.

**This PDF includes:**

Supplementary Notes S1 to S7

Supplementary Tables S1 to S3

Supplementary References

## Supplementary Notes

### 1. The atomistic modeling

#### A. Parameters of the inter-atomic potential for $SrTiO_3$ in the shell model

| | Core charge | Shell charge | $k_2$ | $k_4$ |
|---|---|---|---|---|
| Sr | 3.8847 | -2.0889 | 5580.2 | 0.0 |
| Ti | 6.0887 | -3.2039 | 733.3 | 5500.0 |
| O | 0.8633 | -2.4235 | 22.7 | 5324.0 |
| Short range | *A* | $r_0$ | *C* | |
| Sr-O | 548.201 | 0.38141 | 0.0 | |
| Ti-O | 2173.426 | 0.27463 | 4.39 | |
| O-O | 1769.965 | 0.32975 | 256.92 | |

**Table S1** Parameters of the inter-atomic potential for $SrTiO_3$ (units are electron for charges, Å for distances and eV for energy). The short-range interaction takes the form of a Buckingham potential $A\exp(-r/r_0)+C/r^6$, where $r$ is a distance between a pair of atoms. An anharmonic spring interaction $k_2r^2+k_4r^4$ acts between cores and shells within a given atom.

#### B. Supercell optimization and lattice dynamics

The calculations were done for the $(PbTiO_3)_{16}/(SrTiO_3)_{16}$ system with a supercell comprising 28×1×32 perovskite unit cells. The number of cells in the *x*-direction was taken based on the experimental superlattice peak position indicating the average length scale for a vortex pair. Hence, in all of the atomistic calculations, one such pair is treated with periodic boundary conditions in all directions. In-plane strain was imposed with the strain values taken with respect to the $PbTiO_3$ cubic lattice parameter calculated with the Sepliarsky and Cohen potential at 0 K (3.8845 Å). Out-of-plane supercell length was allowed to change during the structural optimization involving a rational function optimizer. For a given strain value the optimization was followed with the calculation of dynamical properties including phonon/vortexon mode frequencies, eigenvectors (both from the diagonalization of the dynamical matrix), and frequency-dependent permittivity. The latter was taken as proportional to the oscillator strengths of modes. Local polarization maps presented in Figs. 1-3 and Extended Data Fig. 4 were calculated as dipole moments of Ti-centered perovskite unit cells. The amplitudes of the atomic displacement in these maps are enlarged by a factor of 2.5 for better visualization.

The collective circular pattern of atomic displacements carries a mechanical angular momentum within each vortexon supercell. The magnitude of this angular momentum carried by atoms in a volume of 14×1×16 unit cell can be estimated by the measured ionic displacements (Fig.S8e) and frequency (0.08 THz) using $L = \sum_i mR_i \times d\,R_i/dt$, where $\boldsymbol{R}$ is the vector from the vortexon core to the *i*th atom in the supercell of vortexon (Extended Data Fig. 4a). Taking lead ionic displacement as an example, the maximum of this angular momentum at the peak of the oscillating period can reach ~$4.4 \times 10^{-30} \mathrm{m^2 kg s^{-1}}$.

The atomistic model shows that an interfacial $SrTiO_3$ layer within approximately ~3 unit-cells of the interface is also moderately polarized. Upon the THz excitation, the interfacial unit cells in the $SrTiO_3$ layers couple with the strongly driven $PbTiO_3$ layers and contribute to the overall structural response probed by specular reflections. The THz field does not excite $SrTiO_3$ phonons strongly at the temperature range measured in our experiment, but its interaction with $SrTiO_3$ could become more important at cryogenic temperature[30].

### C. Relations to phason modes in nanotwinned ferroelectrics

It is interesting to compare the observed dynamics with that of nanotwinned $BiFeO_3$ crystals studied in Ref. 18. The most prominent polar mode discussed in Ref. 18 was the inhomogeneous phason mode at 0.3 THz, corresponding to the anti-phase longitudinal oscillations of neighboring 71-degree domain walls. In such an oscillation, domain sizes with one polarity are periodically increased at the expense of those of the other polarity and vice versa. These domain proportion fluctuations cause the pronounced infrared activity of the mode in a direction transverse to the domain wall periodicity. Likewise, the vortexon mode identified in the present study is the most characteristic polar mode of the vortex structure. Similar to the pseudo-phason of the one-dimensional incommensurate modulated structures, it involves anti-phase oscillations of adjacent areas of sizable polarization gradient. The reason for its relatively low frequency is common - the sliding of the areas with polarization gradients in the lattice is realized without any appreciable distortion of their polarization profiles. However, the polarization gradient areas in the vortex structure are percolated and mutually interlinked. As a result, there is no strict analogy of inhomogeneous phason mode in the vortex phase and no strict analogy of vortexon in the lamellar phase, since the spectrum of the vortex phase is much more complicated.

## 2. Phase-field modeling

### A. Material parameters used in the phase-field model

| Coefficient | $PbTiO_3$ | $SrTiO_3$ |
|---|---|---|
| Landau coefficient (Temperature in K) | | |
| $\alpha_1$ ($10^7$ $C^{-2}m^2N$) | $0.038(T - 752)$ | $4.05[\coth(54/T) - 1.056]$ |
| $\alpha_{11}$ ($10^8$ $C^{-4}m^6N$) | −0.73 | 17 |
| $\alpha_{12}$ ($10^9$ $C^{-4}m^6N$) | 0.75 | 4.45 |
| $\alpha_{111}$ ($10^8$ $C^{-6}m^{10}N$) | 2.6 | |
| $\alpha_{112}$ ($10^8$ $C^{-6}m^{10}N$) | 6.1 | |
| $\alpha_{123}$ ($10^9$ $C^{-6}m^{10}N$) | −3.7 | |
| Gradient energy coefficient | | |
| $G_{11}$ ($10^{-10}$ $C^{-2}m^4N$) | 1.038 | 1.038 |
| $G_{12}$ ($10^{-10}$ $C^{-2}m^4N$) | 0 | 0 |
| $G_{44}$ ($10^{-10}$ $C^{-2}m^4N$) | 0.519 | 0.519 |
| Elastic stiffness | | |
| $c_{11}$ ($10^9$ Pa) | 230 | 330 |
| $c_{12}$ ($10^9$ Pa) | 100 | 100 |
| $c_{44}$ ($10^9$ Pa) | 70 | 125 |
| Electrostrictive coupling coefficient | | |
| $q_{11}$ ($10^{10}$ $C^{-4}m^2N$) | 1.142 | 1.26 |
| $q_{12}$ ($10^9$ $C^{-4}m^2N$) | 0.49 | −13 |
| $q_{44}$ ($10^9$ $C^{-4}m^2N$) | 3.718 | 11.96 |
| Dielectric permittivity ($\kappa_0 = 8.85\times10^{-12}$ $Fm^{-1}$) | | |
| $\kappa_b$ ($\kappa_0$) | 10 | 10 |

**Table S2** The parameters of $PbTiO_3/SrTiO_3$ in the phase-field model

## B. Derivation of the effective polarization mass density $\mu$

The key parameter allowing to determine the dynamics is the effective polarization mass density $\mu$. Within the LGD theory of ferroelectrics, polarization order parameter is intimately connected with the soft mode mechanism of phase transitions. For PTO, ferroelectric polarization originates from the condensation of a transverse optic mode (Last mode) at the Curie temperature and this mode subsequently recovers with further cooling. Thus, the effective polarization mass density $\mu$ is determined by the temperature-dependent dynamics of the soft mode and bulk dielectric response[48]. Here we derive the $\mu$'s using a simplified formula, considering $P$ in Eq. 2 as a scalar and taking its derivative with respect to $P$:

$$\mu\ddot{P} + \gamma\dot{P} = -2\alpha_1 P - 4\alpha_{11} P^3 - 6\alpha_{111} P^5 \qquad \text{(Eq. S5)}$$

which is an anharmonic oscillator equation. Its eigenfrequency corresponds to the zone-center soft phonon energy for stress-free bulk PTO or STO phases at a certain temperature, values of which are well documented in previous inelastic neutron and light scattering studies[49–51]. Due to uncertainties in those experimentally derived coefficients and the fact that PTO/STO in the superlattice films could be quite different from the bulk phases, the obtained $\mu$'s served as initial input for the phase-field model where P is tensorial and with the full terms in free energy in Eq. (1) and were tweaked within one order of magnitude to best reproduce the observed modes. The optimized parameters are $7.53 \times 10^{-17}\mathrm{kg \cdot m^3A^{-2}s^{-2}}$and $6.89 \times 10^{-17}\mathrm{kg \cdot m^3A^{-2}s^{-2}}$ for PTO and STO, respectively. Using a smaller effective polarization mass density $6.27 \times 10^{-18}\mathrm{kg \cdot m^3A^{-2}s^{-2}}$, the frequencies of bulk PTO E(TO) and $A_1$(TO) modes are reproduced, validating the dynamical phase-field simulation. The $\gamma$'s were estimated from the phonon widths and also optimized against the experiment, which yields $2\times 10^{-7}\Omega \cdot m$.

It is expected that the effective polarization mass density in the investigated superlattices can be different from the bulk values. The modification of tetragonality of the PTO unit cell and complex polarization coupling in the vortex structure can lead to different dynamical properties of the PTO unit cell in the superlattice. For example, according to the atomistic model described above, the soft mode eigenvector components of the bulk PTO homogeneously strained to the same average strain as that of the superlattice yield polarization mass densities $3.73 \times 10^{-18}\mathrm{kg} \cdot \mathrm{m}^3\mathrm{A}^{-2}\mathrm{s}^{-2}$, $4.08 \times 10^{-18}\mathrm{kg} \cdot \mathrm{m}^3\mathrm{A}^{-2}\mathrm{s}^{-2}$ and $7.67 \times 10^{-18}\mathrm{kg} \cdot \mathrm{m}^3\mathrm{A}^{-2}\mathrm{s}^{-2}$, for components along the *x*, *y*, and *z* axes, respectively, in agreement with the parameter used in the phase-field simulation within one order of magnitude. In parallel, both phase-field and atomic model calculations predict a considerable decrease of the frequency of PTO E(TO) mode in the investigated superlattices. To verify this theoretical prediction, we performed THz absorption spectroscopy measurements (Supplementary Note S7) in the same type of samples as those used in the x-ray studies. We found a prominent absorption peak around 2 THz, which is lower than the lowest frequency mode in the bulk PTO crystal and agrees well with the predicted modified frequency of PTO E(TO) mode in the superlattice structure.

### C. Estimation of x-ray scattering intensity

A vortex configuration generated by the phase-field model was represented by a set of continuum field variables, and there must be a corresponding large-box atomistic structure based on which x-ray structure factors can be evaluated. The Cauchy–Born rule is generally assumed to bridge the continuum and atomistic models. It states that, in the homogenous strain limit, the atomic motions of a crystal lattice follow its macroscopic displacement field[52,53]. While this rule has not been *a priori* validated for the PTO/STO superlattice system, remarkably close matching between the atomically resolved images and phase-field models revealed in the previous studies[5,33] suggests the soundness of such an approximation.

Naturally, we chose the cubic phase of PTO/STO as the reference configuration in accordance with the phase-field model; the deformed configurations were the simulated temporal evolution of the vortices upon THz excitation. We first derived the Pb ion coordinates according to the elastic displacement field sampled at the cubic lattice grid points. Here a nearest-neighbor smoothing method was adopted, following the spirit of Zimmerman *et al* [52], to minimize errors in mapping continuum field variables onto discreet lattice vectors. We then filled Ti ions into the center-of-mass of each (non-cubic) eight-Pb ion cell offset by a displacement vector linearly scaled with the local polarization field. Note that the resultant Pb sublattice already incorporated the information of the polarization field via the contribution of eigenstrains. For simplicity, we excluded the O sublattice in our calculation since the O ions only contribute to a minor scattering intensity (< 20% for bulk PTO) due to its small atomic scattering factor. Despite all these approximations, we found a good agreement between the simulated and measured *H0L* reciprocal space maps around the 004 reflection (see Extended Data Fig. 8), attesting to the overall veracity of our approach. The sensitivity for probing atomic displacement in different vortex modes could also depend on the Miller indices of the probed peaks.

### D. Electric-field tunable vortexon mode.

The phase-field simulations were also used to study the tunability of the vortexon modes. For example, it predicts that the frequency and amplitude of the vortexon can be tuned by applying a static electric field (Extended Data Fig. 6c,d), providing an effective control for high-frequency ferroelectric nanodevices. This tunability is achieved by changing the equilibrium microscopic vortex configurations. The phase-field simulation shows that the applied static electric field can effectively offset the vortex cores to more staggered configurations. This effect is similar to applying an in-plane strain (Fig. 4a), which leads to the hardening of the vortexon mode.

## 3. Analytical description of the dynamics of vortexon

The dynamics of extended objects such as domain walls, domain bubbles, polar vortices, vortexons, etc. can be described by their equations of motion associated with their effective mass and spring constants [4,34,54–61]. For the particular example of polar vortices and mechanical vortexons in the $PbTiO_3/SrTiO_3$ superlattice, we describe their dynamics by modeling the vertical displacement $\xi$ of the vortex cores away from the center plane of the $PbTiO_3$ layer (Fig. 3**c**). The kinetic energy of the vortex motion is written as $\frac{1}{2}K\dot{\xi}^2$, where $K$ is the effective mass of the vortex and dot represents time derivative. The potential energy is $\frac{1}{2}B_2\xi^2 + \frac{1}{4}B_4\xi^4$, where $B_2$ and $B_4$ are the first-order and third-order effective spring constants, respectively. The non-zero $B_4$ term reflects the anharmonicity of the potential surface. We found that $B_2$ is linearly dependent on the strain $\varepsilon$ following $B_2 = -b_2(\varepsilon - \varepsilon_A)$ with $\varepsilon_A = -0.17\%$ and$b_2 = 7.5 \times 10^{10}\text{kg m}^{-1}\text{s}^{-2}$ at room temperature, while $B_4 = 3.0 \times 10^{26}$ kg $\text{m}^{-3}s^{-2}$ and $K = 6.5 \times 10^{-16}kg\ m^{-1}$ are strain-independent[54].

The equation of motion for the polar vortex core can be written as

$$\ddot{\xi} = -\frac{B_2}{K}\xi - \frac{B_4}{K}\xi^3 \qquad \text{(Eq. S6)}$$

Under a strain $\varepsilon < \varepsilon_A$, $B_2 > 0$, the potential energy has a single minimum at $\xi = 0$, corresponding to the symmetric vortices under compressive strains found in atomistic and phase-field simulations. Under a linear approximation, Eq. S6 yields a harmonic oscillator with a frequency of

$$\omega = \sqrt{\frac{B_2}{K}} = \sqrt{\frac{b_2(\varepsilon_A - \varepsilon)}{K}},\ \varepsilon < \varepsilon_A \qquad \text{(Eq. S7)}$$

For $\varepsilon > \varepsilon_A$, $B_2 < 0$, the potential energy changes to a double-well function with two minima at $\xi = \pm\sqrt{-\frac{B_2}{B_4}}$, corresponding to the staggered vortices under tensile strains found in atomistic and phase-field simulations. Consequently, the motion of the vortices described by Eq. S6 becomes a harmonic oscillator with equilibrium positions of $\pm\sqrt{-\frac{B_2}{B_4}}$ and a frequency of

$$\omega = \sqrt{\frac{-2B_2}{K}} = \sqrt{\frac{2b_2(\varepsilon-\varepsilon_A)}{K}},\ \varepsilon > \varepsilon_A \qquad \text{(Eq. S8)}$$

under a linear approximation.

## 4. Modeling acoustic phonon propagation in superlattices

Superlattice acoustic modes refer to a specific set of lattice vibrations as acoustic phonons propagate and scatter across the alternating layers of the superlattice, which are governed by superlattice geometry and mechanical properties of materials. The Brillouin zones are folded according to the periodicity of the superlattice. The energy gaps open at the centers and edges of the folded Brillouin zones. The phonon dispersion relation of the superlattice is described by[38]

$$\cos(q(d_A + d_B)) = \cos(\omega d_A/v_A)\cos(\omega d_B/v_B) - \frac{1+\delta^2}{2\delta}\sin(\omega d_A/v_A)\sin(\omega d_B/v_B) \qquad \text{(Eq. S9)}$$

where $\omega$, $q$ are the frequency and wave vector of the acoustic phonons, $\delta = \rho_A v_A/\rho_B v_B$, and $d$, $v$, $\rho$ are the thickness, sound velocity, and mass density of A or B layers. In our case, A and B denote $PbTiO_3$ and $SrTiO_3$, respectively. Experimentally, upon 400 nm optical excitation, we measured the diffraction intensity oscillation of the first-order diffraction peak of the superlattice (Extended Data Fig. 9a). The Fourier transform reveals the fast frequency components at 0.43 and 0.56 THz (Extended Data Fig. 9b). No significant responses were observed on the satellite diffraction peaks. Using the parameters listed in Table S3, the dispersion curves of longitudinal and transverse acoustic phonon branches were calculated and plotted in Extended Data Fig. 9c.

The measured frequencies (solid circles) agree with the predicted superlattice acoustic phonon modes (crossings of phonon dispersion curve with the probed scattering vector), showing a nominal acoustic response upon optical excitation. However, the set of modes (stars) as revealed upon the THz-field excitation were not predicted by this model, indicating new modes that arise from the unique configuration of polar vortices in the sample.

| | $d$ (nm) | $\upsilon$ (km/s) | $\rho$ (g/cm$^3$) |
|---|---|---|---|
| $PbTiO_3$ | 6.4 | Longitudinal: 4.2 [62] | 7.52 |
| | | Transverse: 3.0 [62] | |
| $SrTiO_3$ | 6.4 | Longitudinal: 7.8 [63] | 4.81 |
| | | Transverse: 4.4 [64] | |

**Table S3** Parameters for modeling acoustic phonon in the superlattice

In addition to the intensity modulation of the superlattice diffraction peak, the overall peak shift also occurs as the strain wave propagates through the whole thickness of the superlattice film of $D$ =100 nm, and results in a large amplitude change of the measured diffraction intensity at fixed x-ray incident angle (Extended Data Fig. 9a). This slow but large intensity modulation is related to the Bragg peak shift and its zero-crossing at 18.6 ps indicates the strain wave completes a single pass of the film. This is in agreement with the calculated time scale $t = \frac{D}{\upsilon_s}$ =18.6 ps, where the averaged sound velocity $\upsilon_s$ of the superlattice is determined by $\frac{d_A+d_B}{\upsilon_s} = \frac{d_A}{\upsilon_A} + \frac{d_B}{\upsilon_B}$.

## 5. Anisotropic structural response and analysis of the vortexon mode

The relative orientation of the THz electric field with respect to the crystalline axes does not affect the frequencies but the amplitude of the observed modes. Both atomistic calculations and phase-field modeling predict that the sample responds stronger to the THz field applied along the $x$-axis. For example, the oscillator strength as obtained in the atomistic calculations is dominant for $xx$ modes (Extended Data Fig. 3a). The amplitude of the vortexon mode as obtained in the phase-field modeling is the largest when the THz field is applied along the $x$-axis (Extended Data Fig. 6a,b). Experimentally, we found that the response of the 004 Bragg peak was about a factor of two stronger when the THz field has a projection along the $y$-axis, but not the $x$-axis as the theory predicted (Extended Data Fig. 5). This discrepancy can be partially addressed by substantial non-zero cross-terms of permittivity ($\varepsilon_{xy}$) which are revealed by the atomistic model (Extended Data Fig. 3b). Therefore, the THz field along the $x$-axis can also excite a response along the $y$-axis (axial direction of vortex), which points to a three-dimensionally coupled response of these collective modes. It remains challenging to accurately predict the dynamics along the $y$-axis since the simulation size along the $y$-axis was only one unit cell thick, which was much smaller than the actual length of the vortex along the $y$-axis (~ µm).

To examine the frequency and phase of the diffraction intensity oscillations of vortexon mode observed in 004 vortex superlattice peak, we subtract a baseline variation shown in Extended Data Fig. 5b and reveal that the oscillation period slightly shortens at longer delays, as indicated by the shift of the minima of the measured data compared with a sinusoidal wave. A possible cause is the strain wave propagation across the film that exerts a transient strain and thus dynamically tunes the mode frequency. We also note that the fitting of $\sin[\omega(t - t_0) + \psi\,]$ to the background-subtracted oscillation yields $\psi$ =10±5 degrees (Extended Data Fig. 5b). The offset $t_0$=1 ps was taken into account because the THz field ramps up at 1 ps before the nominal time zero as defined by the arrival time of the THz peak field. The diffraction intensity change thus is close to being described by a sinusoidal oscillation, which is consistent with the oscillations measured at 023 and 113 peaks (Fig. 2a) and indicates an impulsive rather than displacive excitation[65] by the THz field.

## 6. Simplifications of the theoretical modeling

The atomistic model and dynamical phase-field simulation aim at understanding the origin of the observed modes and have reproduced most of the observed phenomena as quantitatively as possible, including the frequencies of eigenmodes, the resultant structural dynamics as probed by femtosecond x-ray diffraction, and their strain dependence. But some experimental observations do not yet agree with the theory including the polarization dependence of the response and the dynamical changes of the x-ray diffractions for high-frequency modes. Two simplifications of the theoretical model comparing with the realistic superlattice samples may help understand these discrepancies. First, the theoretical model is only one-unit-cell thick along the y-axis, which prohibits modeling any collective axial polarization modulation over many unit cells. Second, the theoretical model only contains pure vortex structures, while the real samples contain both vortex and FE structures with a stripe domain size of ~200 nm. Even in the phase-field simulation which comprises the full thickness of the vortices, the simulation size is only $10^{-8}$ of the experimentally probed volume. The coexistence of FE and vortex nanodomains and their interactions across domain boundaries may play roles that were not accounted for in the theoretical models. These simplifications may be improved and refined by considering a full-size 3D model.

## 7. THz spectroscopy measurements

To further support the observation of sub-THz collective modes, we performed THz absorption spectroscopy measurements. The measurements were performed in a dry nitrogen-purged THz time-domain spectroscopy (THz-TDS) system at room temperature[66,67]. The 100 fs laser pulses at the central wavelength of 800 nm were generated by Ti: Sapphire laser systems. The THz emission is based on the optical rectification of $\langle 110 \rangle$-cut ZnTe crystal. Separately, low-temperature grown GaAs based photoconductive antenna was also used for THz generation driven by a Ti: Sapphire oscillator. The generated THz field was collimated and focused by a pair of parabolic mirrors onto the samples. The transmitted THz waveform is analyzed by the electro-optical sampling technique. To obtain the THz dielectric function and minimize the contributions of the substrate, a bare substrate without superlattice film was used as the reference. The time-domain from sample and reference measurements were Fourier transformed to obtain the frequency domain spectra $E_{sam}(\omega)$ and $E_{ref}(\omega)$. The transmission coefficient was computed as $\tilde{t}(\omega) = E_{sam}(\omega)/E_{ref}(\omega)$. The complex dielectric function was extracted from the transmission coefficient[68].

First, we measured the THz transmission of the superlattices grown on $DyScO_3$ (DSO) substrate (bottom panel, Extended Data Fig. 7d), the same type of samples used in the x-ray measurements. However, due to the coexistence of vortex and FE domain structures and a strong substrate attenuation that could obscure the absorption feature below 0.5 THz, we did not see clear absorption peaks reliably in this frequency range (large error bars in the shaded regime). At higher frequency range, we have observed an absorption feature around 2 THz, which is absent in the STO and PTO films at room temperature. This result is in agreement with the modified PTO E mode as the dynamical phase-field simulation predicted in Supplementary Note S2.

We further studied the PTO/STO superlattice grown on $Sr_2Al_{0.3}Ga_{0.7}TaO_6$ (SAGT) substrate in order to avoid the complications associated with the DSO substrate. The absorption at 0.2- 0.5 THz range in SAGT is smaller than DSO and the superlattice only contains vortex structures without FE domains. The x-ray diffraction measurements showed the existence of vortex satellite peaks that are consistent with the vortex structures (Extended Data Fig. 7c). In this sample, we observed an absorption peak at 0.32 THz at room temperature (Extended Data Fig. 7d), further support that these modes are uniquely related to the polar vortex structure.

It is worth noting that these THz measurements are not straightforward due to the complications of phase mixtures and substrate absorption. In particular, the 0.08 THz vortexon mode was not accessible by conventional THz spectroscopy. These characterization challenges further illustrate that the present ultrafast x-ray diffraction experiment is superior in detecting these low-frequency modes. First, ultrafast x-ray diffraction can probe selectively the signal from vortex phase areas only, avoiding the complication of the mixed domain structures or substrate absorption. Second, it has substantially better sensitivity below 0.2 THz, because the limitations caused by weak THz signal propagation through the sample are circumvented by a completely different principle of detection.

## Supplementary References: